%% file: main.tex
\documentclass{ifacconf}

\usepackage{graphicx}      % include this line if your document contains figures
\usepackage{natbib}        % required for bibliography

\usepackage{color}

\usepackage{amsmath}
\usepackage{amsfonts}
\usepackage{algpseudocode}
\usepackage{algorithm}
\usepackage{graphicx}

\usepackage{tikz}
\usetikzlibrary{positioning,angles,quotes,matrix}
\usepackage{multirow}
\usepackage{multicol}

\usepackage{capt-of}% or \usepackage{caption}
\usepackage{booktabs}
\usepackage{varwidth}

\DeclareMathOperator*{\argmax}{\arg\!\max}
\DeclareMathOperator*{\argmin}{\arg\!\min}

\newtheorem{remark}{\bf Remark}[section]
\begin{document}
% Color codes
\newcommand{\red}[1]{{\color{red}  #1}}
\newcommand{\blue}[1]{{\color{blue}  #1}}
\newcommand{\green}[1]{{\color{green}  #1}}

\newcommand{\oc}[1]{\textbf{O}_{#1}}

\begin{frontmatter}

\title{Optimal Policy Synthesis from A Sequence of Goal Sets with An Application to Electric Distribution System Restoration} 

\author[First]{\.{I}lker I\c{s}{\i}k} 
\author[First]{Onur Yigit Arpali} 
\author[First]{Ebru Aydin Gol}

\address[First]{Department of Computer Engineering, Middle East Technical University,
   Ankara, Turkey (e-mail: isik.ilker\_01,arpalion,ebrugol@metu.edu.tr).}

\begin{abstract}                % Abstract of not more than 250 words.
Motivated by the post-disaster distribution system restoration problem, in this paper, we study the problem of synthesizing  the optimal policy for a Markov Decision Process (MDP) from a sequence of goal sets. For each goal set, our aim is to both maximize the probability to reach and minimize the expected time to reach the goal set. The order of the goal sets represents their priority. In particular, our aim is to generate a policy that is optimal with respect to the first goal set, and it is optimal with respect to the second goal set among the policies that are optimal with respect to the first goal set and so on. To synthesize such a policy, we iteratively filter the applicable actions according to the goal sets. We illustrate the developed method over sample distribution systems and disaster scenarios.
 % Each goal set induce an optimization objective: minimize the expected time to reach the goal set (over the paths that reach the goal set) among the policies that maximize the probability to reach. The order of the goal sets from the given sequence represents their priority. In particular, the resulting policy is optimal w.r.to the first goal set, and it is optimal w.r.to the second set among the policies that are optimal with respect to the first set and so on.  
 % % For example, for a goal sequence of size two, our aim is to synthesize a policy that is optimal for the second goal set among the policies that is optimal for the first goal set.
 % is optimal for the first goal set, and among all the policies optimal for the first 
% For a goal set G, the optimal policy minimizes the expected number of steps to reach G over the paths that reach G among all the policies that maximize the probability to reach G. 
% Thus, the optimal policy for the goal sequence satisfies this objective for each goal set in the given order. 
\end{abstract}
\begin{keyword}
Stochastic systems, Energy and power networks, Specification
% SISTEMDE UP-TO 3 seciliyor.
% Five to ten keywords, preferably chosen from the IFAC keyword list.
\end{keyword}

\end{frontmatter}
%===============================================================================

\section{Introduction}

\red{
% \begin{itemize}
%     \item Short paragraph on MDP. 
%     \item Value iteration description --- related works. 
%     \item Related works on the policy synthesis for our specification: sequence of target/goal sets. (Is there any work on same/similar specification?)
%     \item Dead-end paper, their motivation and results. 
%     \item What we do -  / contributions
%     \item A short summary of our power system restoration paper, and description of how we apply the developed method to this particular MDP.
% \end{itemize}
}
\input{Sections/intro}
\section{Preliminaries and Problem Formulation}\label{sec:prelim}
\input{Sections/prelim}

\section{Policy Synthesis}\label{sec:synthesis}
\input{Sections/policy_synthesis}

\section{Synthesis for Distribution System Restoration}\label{sec:restoration}
% \red{
% \begin{itemize}
%     \item How we generate MDP (very short), what a state represents
%     \item Target set : max/min
%     \item Target set : min/min 
%     \item Results on sample systems
% \end{itemize}
% }
\input{Sections/mdp_construct}
\input{Sections/maxmin}

\input{Sections/example/main}

\section{Conclusion}\label{sec:conc}
In this paper, we develop a method to synthesize a policy for an MDP from a sequence of goal sets. % Each goal set induce an optimization objective: minimize the expected time to reach the goal set (over the paths that reach the goal set) among the policies that maximize the probability to reach. The order of the goal sets from the given sequence represents their priority. In particular, the resulting policy is optimal w.r.to the first goal set, and it is optimal w.r.to the second set among the policies that are optimal with respect to the first set and so on.  
Our method is based on iterative filtering of the applicable actions, which yields fast performance. While we focus on minimization of the expected time to reach the goal set, it is straightforward to generalize this approach to other optimization criteria. 
We show that the developed method allows us to synthesize a restoration strategy with prioritized components for an earthquake damaged distribution system.
Although this application motivated our study, the algorithm can be applied to any MDP. Future research directions include selection of goal sets to keep the value function within a predefined range as in~\citep{remi}.

% \red{
% From template: "A conclusion section is not required. Although a conclusion may review the main points of the paper, do not replicate the abstract as the conclusion. A conclusion might elaborate on the importance of the work or suggest applications and extensions."

% Importance is covered but can be elaborated.
% \begin{itemize}     \item  % To best of our knowledge, are we the first one to optimize according to a sequence of goal sets, unlike only one goal set (\cite{s3p}) or selecting safest goals (\cite{remi})?
% \item Filtering actions yields faster performance in practice, and it is easy to implement. Can be generalized to other optimization criteria easily.
% \item Despite the specific application, our prioritization method can be applied to any MDP (Bunu bence de conclusion da belirtebiliriz.) \end{itemize}

% Potential Extensions:
% \begin{itemize}
%    \item Minimizing expected time among policies that minimize value, i.e., the last optimization criterion presented in this paper becomes the first
%    \item Minimize the expected time only if the value function is within an allowed bound, don't let value function increase too much while optimizing other criteria
%    \item Application to other areas where MDPs are used to model problems
%\end{itemize}
% }

% Referans eklenecek (Ebru)

%Tablo 9daki iki senaryo icin bizim policy'nin ve Arpali policy'sinin Arpali value fonksiyonuna gore degeri  (Ilker, Onur)

%Son tablo icin text yazilmasi / olursa 3 bus icin durum. (Ilker)

% Conclusion (Ebru)

% Bastan sona tekrar okunmasi (All)

%\bibliographystyle{IEEEtran}
\bibliography{bib.bib}

\end{document}

%% file: Sections/intro.tex
The rapid restoration of electricity after an earthquake is essential in disaster management~\citep{Yuan2016,Qiu2017}. After an earthquake, a complete blackout may be experienced in the distribution system, i.e., all breakers open for safety considerations. During the restoration process, starting from the energy resources (transmission grid or distributed energy resources) energization actions are applied iteratively, and the aim is to energize the whole system as soon as possible. Black-start is already a hard problem since electrical and topological constraints have to be considered. This problem becomes even more complex after an earthquake as the field instruments may be damaged. The system operator, therefore, needs a decision support tool to guide the restoration process. 

\cite{Gol2019} model the restoration process as a Markov Decision Process (MDP) by using the probability of failure (i.e. destruction) values of the field instruments, and provide a restoration strategy. These values are computed by using the peak ground acceleration values recorded during the earthquake~\citep{seismic}. The MDP model reduces the synthesis of an optimal restoration strategy problem into an MDP policy synthesis problem.  \cite{Gol2019} define the objective as the minimization of the overall restoration time that maps to a stochastic shortest path (SSP) problem~\citep{ssp}. The SSP formulation results in policies with non-optimal average restoration time over the system components, i.e. buses supplying electricity to buildings/customers. This issue is addressed by~\cite{Arpali2019}. They define the state cost as the number of unenergized components, which results in minimizing the average restoration time. In  either of the cases, a prioritization over the system components is not possible. However, in a post disaster scenario, rapid restoration of electricity for some of the components can be more important than others. For example, energization of hospitals within the area affected by the earthquake or energizing the base stations in areas with collapsed buildings can be prioritized. In the MDP model from~\cite{Gol2019}, prioritization for a target set maps to minimization of the expected time to reach a set of MDP states, which also maps to SSP. However, in general, in goal-oriented policy synthesis problems including SSP, an optimal policy is synthesized under the assumption that the probability to reach the goal set is $1$. However, this assumption might not hold for the considered goal sets. Furthermore, it is not straightforward to integrate a sequence of a goal sets in this formulation.

\cite{s3p} introduces a theoretical framework that primarily maximizes the probability to reach a goal set, and 
 then minimizes the expected cost only over the paths that reach a goal. Thus, the cost optimization is only performed over the policies that achieve the optimal probability to reach the goal set, that is not necessarily $1$. \cite{remi} extend this approach to select a subset of goal states and synthesize a policy to visit each of these selected goal states. However, a prioritization among these sets is not considered. 
 
 %  among the policies with the maximal probability to reach $G_i$
In this work, we extend the method developed by~\cite{s3p} to synthesize a policy from a sequence of goal sets $\textbf{G} = [G_1, G_2, \ldots, G_n]$. Here, we synthesize the optimal policy $\pi$ that achieves the objectives in the given order and finally minimizes the  value function of the MDP. Essentially, each goal set introduces two objectives: maximization of the probability to reach $G_i$ and minimization of the expected number of steps to reach $G_i$ over the set of paths that reach $G_i$. In order to find the optimal policy satisfying the objectives in the given order, we iteratively filter  actions that do not attain the optimal values for each goal set and finally synthesize a policy minimizing the value function with respect to the remaining actions. Thus, we ensure that the resulting policy is optimal for the goal set $G_1$, and it is optimal for the goal set $G_2$ among the policies that are optimal for $G_1$ and so on. We show that the developed method allows us to prioritize among the different field components in the restoration problem. We illustrate our results on two examples, and compare the results with the optimal strategies obtained in~\citep{Gol2019,Arpali2019}.

The policy synthesis problem is also studied under temporal logic specifications~\citep{Baier2004,ltl3}. \cite{MortezaPCTL} synthesize a policy satisfying a probabilistic computation tree logic (PCTL) formula. Both \cite{ltl2} and
\cite{6702421} primarily maximize the probability of satisfaction of a linear temporal logic (LTL) formula then minimize the considered cost. 
% minimize the expected cost between visits of a set of states over the policies that maximize the probability of satisfying a linear temporal logic (LTL) formula. 
Sequential tasks can easily be specified in temporal logics such as a strict order between the tasks (e.g. satisfy A and then B) or a set of tasks without any particular order (e.g. eventually A and eventually B). However, a prioritization among the tasks without a strict order can not be directly integrated to an LTL formula. 

The rest of the paper is organized as follows. The preliminary information and the problem formulation are given in Sec.~\ref{sec:prelim}. The proposed policy synthesis method is explained in Sec.~\ref{sec:synthesis}. The application to the electric distribution system restoration problem and the results over sample systems are given in Sec.~\ref{sec:restoration}. Finally, the paper is concluded in Sec.~\ref{sec:conc}.

%% file: Sections/prelim.tex
% \item MDP definition
% \item Cost definition
% \item Definition of the specification: Sequence of states + optimality criteria
% \item Problem formulation

%  goal-oriented , and $G \subset S$ is a set of goal states

\subsection{Markov Decision Process}\label{sec:MDP}

A Markov Decision Process is a tuple $M = (S, A, T, c)$ where $S$ is a finite set of states, $A$ is a finite set of actions,  $T: S \times A \times S \rightarrow [0,1]$ is a transition probability function, i.e., $T(s,a,s')$ is the probability of transitioning to state $s' \in S$ by taking action $a \in A$ from state $s \in S$, and $c: S \times A \times S \rightarrow \mathbb{R}$ is a cost function that defines the cost $c(s,a,s')$ incurred when $s'$ is reached from $s$ by applying action $a$ \citep{bert}. 
% EBRU:  kalani incomplete ve gereksiz -- where $T(s,a,s) = 1$ and the cost is $0$ for each $s\in G$.
%If the cost function does not depend on the taken action and the resulting state, as it is the case with distribution grid problem, it is shortly defined as $c: S \rightarrow \mathbb{R}$.  
% Cost determines the favourability of a state/action.
The function $app: S \rightarrow 2^{A}$ is used to denote the set of applicable actions for a state, i.e., $\sum_{s'\in S} T(s,a,s') = 0$ if $a \not \in app(s)$, where $2^A$ is the power set of $A$. 
% Besides, we also use the function $app: S \rightarrow 2^{S}$ mapping a state $s \in S$ to the set of applicable actions in that state.

A deterministic stationary policy $\pi: S \to A$ assigns an action $\pi(s) \in A$ to each state $s \in S$. Given a policy $\pi$, we define an n-step value  function $V_n^{\pi}$ that represents the discounted sum of the expected cost incurred by following the policy $\pi$ for n-steps. For a given discount factor $\gamma \in (0,1] $, the value function is defined as: 
% For a discount factor $\gamma \in (0,1]$,
% Typically, an MDP aims to find the optimal policy $\pi^*$ that includes the state-action pairs which minimize the cumulative cost, computed by a value function $V: S \rightarrow \mathbb{R}$.
% For a given discount constant $\gamma$ within n step, the value function in (\ref{eq:value}) is defined as
\begin{equation}\label{eq:value}
    \begin{aligned}
        V_n^{\pi}(s) &= \sum_{s' \in s} T(s, \pi(s), s') ( c(s, \pi(s), s') + \gamma V_{n-1}^{\pi}(s') ) \\
        V_0^{\pi}(s) &= 0
    \end{aligned}
\end{equation}
% When the discount factor $\gamma$ is $1$, 
% Note that the value function represents the expected cost over $n$-steps.
%The function in \ref{eq:value} is the conventional definition of the value function. Depending on the problem, a value function can be a discounted sum over an infinite horizon, or it can be a undiscounted sum over a finite horizon. 

\subsection{Optimization Criteria}
\label{sub:optimization_criteria}

% \blue{Optimization kriterlerini formulasyon olarak tanimlayip daha acik ve net sekilde anlatmak. Onceki paperdaki constraint tanimlari gibi yapabiliriz. Aksiyonlar state lere gore tanimlanip, filtreleme sonucu olusan aksiyon setleri belirtelim. Action setleri state tabanli yazmak icin 'app'(stateden a setinin subsetine giden bir fonksiyon) gibi bir fonksiyon tanimlayalim}\\

In goal-oriented policy synthesis problems, the optimization criteria, in general, is defined as either maximizing the probability to reach the goal set or minimizing the value function among the policies that reach the goal set with probability $1$~\citep{bert,Kolobov2011}. In~\cite{s3p}, the two objectives are combined to generate a policy from the set of policies that maximize the probability of to reach the goal set, such that it minimizes the cost of the paths that reach the goal set. In this work, we extend this approach to a sequence of goal sets. Here, we recall the probability to reach a goal set in a given number of steps ($P_n^{G,\pi}$) and the accumulated cost averaged over the paths that reach $G$ ($C_n^{G, \pi}$)
definitions from~\cite{s3p}. %, and then formulate the extended problem. % and the expected number of steps to reach a goal set
 % \red{(NOTE: \cite{s3p} does not define expected number of steps to reach the goal; it defines expected cost while reaching the goal)}

Let $G \subset S$ be a set of absorbing goal states, i.e. $T(s,a,s') = 0$ for any $s \in G$, $a\in A$, $s' \in S \setminus G$. $P_n^{G, \pi}(s)$ denotes the probability of reaching the goal set $G \subseteq S$ within $n \in \mathbb{N}$ steps by executing policy $\pi : S\to A$ from state $s \in S$. $P_n^{G, \pi}(s)$ is computed as:

\begin{align}
\label{eq:pr_Gpin}
	P_n^{G, \pi}(s) &= \sum_{s' \in S} T(s, \pi(s), s') P_{n-1}^{G, \pi}(s'), \\
	P_0^{G, \pi}(s) &=
	\begin{cases}
		0 & \textrm{ if } s \in S \setminus G \\
		1 & \textrm{ if } s \in G
	\end{cases} \nonumber
\end{align}
The expected cost of paths that reach the goal set within $n$ steps is defined for states $s \in S$ with $P_n^{G, \pi}(s) > 0$ as follows (Theorem 1 in~\cite{s3p}):
\begin{align}\label{eq:c_Gpin}
    C_n^{G, \pi}(s) = \frac{1}{P_n^{G, \pi}(s)} \sum_{s' \in S} & T(s,\pi(s), s') P_{n-1}^{G,\pi} {(s')} \times \nonumber \\
    &[c(s,\pi(s),s') + C_{n-1}^{G, \pi}(s')] 
\end{align}
with $C_0^{G,\pi}(s) = 0$ for each $s\in S$.

For the infinite horizon formulation, i.e., when $n$ tends to infinity in~\eqref{eq:pr_Gpin} and~\eqref{eq:c_Gpin}, there exists an optimal stationary policy $\pi^\star$ that minimizes the infinite horizon cost to go function $C_\infty^{G, \pi^\star}$ among all policies that maximize the infinite horizon probability to reach $G$, $P_\infty^{G, \pi^\star}$~\citep{s3p}. Furthermore, the following iterative computation converges to the probability to reach the goal set $G$ under the optimal policy ($P_{\infty}^{*}$ converge to $P_\infty^{G, \pi^\star}$):
\begin{align}
\label{eq:fprgo}
	P_{n}^{*}(s) &= \max_{a \in app(s)} \sum_{s' \in S} T(s, a, s') P_{n-1}^{*}(s'),\\
	P_0^{*}(s) &= 0 \text{ for } s \in S \setminus G, P_{0}^{*}(s) = 1 \text{ for } s \in G. \nonumber
\end{align}
In addition, if the costs of transitions leaving $S\setminus G$ are strictly positive, the following iterative computation converge to the infinite horizon cost to go function for $G$ under the optimal policy ($C_\infty^{\star}$ converge to $C_\infty^{G, \pi^\star}$). $C_n^\star(s) = 0$ if $P_{\infty}^{*}(s) = 0$, otherwise:
\begin{align}
	C_{n}^{*}(s) =& 
	\min_{a \in app(s) : \sum_{s' \in S} T(s,a,s')P_{\infty}^{*}(s') = P_{\infty}^{*}(s) }\frac{1}{ P_\infty^{\star}(s) } \times \nonumber \\
	& \sum_{s' \in S} T(s, a, s') P_{\infty}^{\star}(s') (c(s,a,s') + C_{n-1}^{*}(s'))
	\label{eq:Cgo}
\end{align}
with $C_0^\star(s) = 0$ for each $s\in S$. Intuitively, in~\eqref{eq:Cgo}, an action that minimize the average cost is selected among the actions that maximize the probability to reach $G$.

\subsection{Problem Formulation}\label{sec:problem}

% EBRU: Yok. MDP tuple olarak tanimladik, degistirmeyelim. 
% \green{Should $\textbf{G}$ be added to MDP definition $M=(S,A,T,c, \textbf{G})$?}
In this work, we consider an MDP $M=(S,A,T,c)$ and a sequence of goal sets $\textbf{G} = [G_1, \ldots, G_n]$, $G_i \subseteq S$ for each $i = 1, \ldots, n$ such that each goal set $G_i$ is absorbing.  The ordering of the sequence determines the priority of the goal sets, i.e., $G_1$ has the highest priority. The goal sequence $\textbf{G}$ induce $n$ optimization criteria $\oc{1},\ldots,\oc{n}$.
$\oc{i}$ is to synthesize a policy that minimize the expected number of steps to reach $G_i$ averaged over the paths that reach $G_i$ among the policies that maximize the probability to reach $G_i$. 

% EBRU: OK, simplified.

% \red{a bit hard to understand: expected number of steps to reach G of paths..}\green{expected length of the paths that reach... or just the phrase in parenthesis} of the paths that reach $G_i$ (i.e. averaged over the paths that reach $G_i$) among the policies that maximize the probability of reaching $G_i$. 

We aim at synthesizing a policy $\pi^\star$ that achieves the optimization criteria in the given order and finally minimizes the value function, i.e., synthesize a policy $\pi^{\star}$ that minimizes the value function $V^{\pi^{\star}}_{\infty}$~\eqref{eq:value} among the policies that satisfy $\oc{n}$, among the policies that satisfy $\oc{n-1}$ and so on. Thus, our aim is to satisfy each criteria in the given order and then to minimize the value function.

In order to formally define the synthesis problem, we first introduce a new cost function $c_i: S \times A \times S \to \{0, 1\}$ for each goal set $G_i$:
\begin{equation}\label{eq:Gi_cost}
    c_i(s,a,s') = \begin{cases} 0 \quad \text{ if } s \in G_i \\
    1 \quad \text{ it } s \in S \setminus G_i
    \end{cases}
\end{equation}
The infinite horizon cost to go function $C_{\infty}^{G_i,\pi}(s)$~\eqref{eq:c_Gpin} induced by $c_i$ defines the expected length of the paths that end in $G_i$ (i.e. expected number of steps of the paths that reach $G_i$) and originate from $s$ under the policy $\pi$. Consequently, our goal is to synthesize an optimal policy~$\pi^\star$:

% EBRU: boyle kalabilir - cok aynisi olmasina gerek yok. 
% \green{perhaps we should define a probability-maximizing intermediate step $\Pi_i'$, so that it is easier to understand and more similar to the filtering process ( $app_i'$ and $app_i$) }
\begin{align}
    & \quad\quad\quad \quad \pi^\star(s) \in \argmin_{\pi \in \Pi_n} V_\infty^{\pi}(s), \label{eq:optimalPolicy} \\
    & \text{where for each } i=1,\ldots,n  \nonumber \\
  %   & \Pi_i = \{ \pi \mid \pi(s) \in  \label{eq:PolicySets} \\
%     &    \argmin_{\pi': \forall s' \in S, \pi'(s') \in \argmax_{\pi'' \in \Pi_{i-1}} P_\infty^{G_i, \pi''}(s')} C_\infty^{G_i, \pi'}(s) \} \nonumber \\
    & \Pi_i = \{ \pi \mid \pi(s) \in \argmin_{\pi': \forall s' \in S, \pi'(s') \in A_{i}(s')} C_\infty^{G_i, \pi'}(s) \}  \text{ and } \label{eq:PolicySets}  \\
    & \quad A_{i}(s') = \argmax_{\pi'' \in \Pi_{i-1}} P_\infty^{G_i, \pi''}(s') \nonumber
\end{align}
and $\Pi_0$ is the set of all stationary policies,  i.e., $\Pi_0 = \{\pi: S \to A \mid \pi(s) \in app(s) \text{ for each } s \in S \}$.  Thus, intuitively, $\Pi_i$ is the set of policies that satisfy $\oc{1}, \ldots, \oc{i}$ 
in this order and it is recursively defined in~\eqref{eq:PolicySets}. The optimal policy $\pi^*$ is the policy that minimizes the value function among all policies in $\Pi_n$.
Since all of the policies in $\Pi_n$ satisfy all optimization criteria apart from minimizing the value function, selecting the one that minimizes the value function yields the optimal policy.  The sets of policies $\Pi_1,\ldots,\Pi_n$ formalize the optimization goal and structure our synthesis method.

In order to solve the synthesis problem~\eqref{eq:optimalPolicy}, starting from $app_0 = app$, we iteratively solve two optimization problems for each $\oc{i}$ and filter the sets of applicable actions. The first one filters the actions that do not have the optimal probability to reach the goal set $G_i$. Among the remaining actions, the second one filters the actions that do not yield the optimal expected time to reach the goal set $G_i$, i.e., that do not yield the optimal cost to go~\eqref{eq:Cgo} induced by $c_i$. The filters generate $app_i$ such that $app_i(s)\subseteq app_{i-1}(s)$ for each $i=1,\ldots,n$. At the end of the iterative process, we obtain a new function $app_n: S \to 2^A$ representing the applicable actions for each state. This process ensures that any policy $\pi$ synthesized with respect to $app_n$, i.e., $\pi(s) \in app_n(s)$, satisfies the optimization criteria $\oc{1}, \ldots, \oc{n}$ in the given order. Finally, we solve an optimization problem using $app_n$ to minimize the value function~\eqref{eq:value} over the cost function $c$ from $M$. 

At the $i-th$ iteration of the process, for a state $s \in S$, if the probability to reach the goal set $G_i$ is 0, i.e, $P^\star_{\infty,i}(s) = 0$, then no actions will be filtered for $s$, i.e., $app_{i}(s) = app_{i-1}(s)$. Consequently, the goal sequence impose non-strict requirements in the sense that the aim is to maximize the probability to reach $G_i$ and to minimize the expected number of steps to reach $G_i$ (over the paths that reach $G_i$) only when it is possible. Otherwise, i.e., when $P^\star_{\infty,i}(s) = 0$, no restrictions are applied to $s$. Furthermore, while an order of priority over the goal sets is given, it is not necessary to visit the goal sets in the given order, i.e, a path generated under the optimal policy $\pi^\star$
might visit $G_i$ before $G_j$ when $i > j$. Such ``soft'' constraints over the order of the goal sets to visit are encountered in various probabilistic planning scenarios. An example of such a scenario is given in Sec.~\ref{sec:restoration}, where the MDP models the restoration of a distribution system and the goal sets represent customers with various priorities, e.g., hospitals,  base-stations for mobile networks, residential areas.

% \red{ Delete for the short version }
Note that in addition to the sequence of goal sets $\textbf{G}$, we require the optimal policy to minimize the expected number of steps to reach $G_i$ for each goal set $G_i$. Whereas, only a goal set $G$ and the cost function $c$ are considered in~\citep{s3p}.

%% file: Sections/policy_synthesis.tex
In this section, we present our solution for the policy synthesis problem~\eqref{eq:optimalPolicy} for an MDP $M$ and a sequence of goal sets $\textbf{G} = [G_1, \ldots, G_n]$. Central to the proposed method is the iterative filtering of the applicable actions with respect to the sequence of the goal sets. Consequently, any policy generated from the remaining actions satisfies each criteria defined from the goal sets in the given order. Finally, we synthesize the policy that minimizes the value function from the remaining applicable actions. We first define the proposed iterative method (see Alg.~\ref{algo:filter}), and then present the details of each step.

\begin{algorithm}[ht]
  \footnotesize
  \caption{PolicySynthesis(M , app, $\mathbf{G}$)}
  \label{algo:filter}
  \begin{algorithmic}[1]
    \Require 
      $M= (S,A,T,c)$ : is an MDP, $app:S \to 2^A$ is the applicable actions function of $M$,
      $\mathbf{G} = [G_1, \ldots, G_n]$: the sequence of goal sets. 
    \Ensure $\pi^\star$ solves~\eqref{eq:optimalPolicy} for $M$, $app$ and $\textbf{G}$
    %satisfies each optimization criteria $\oc{1}, \ldots, \oc{n}$ induced by $\textbf{G}$ in the given order, and $\pi^\star$ is optimal for~\eqref{eq:value} w.r.to $app_n$.
    \State $app_0 = app$
    \For{$i=1 \ to \ n$}
    \State $app_{i'} = MaximizeProbability(app_{i-1}, G_i)$~\label{line:filter1}
    \State $app_{i} = MinimizeExpectedTime(app_{i'}, G_i)$ \label{line:filter2}
    \EndFor
    \State $\pi^\star = MinimizeV(M, app_n)$\label{line:optimal}
  \end{algorithmic}
  
\end{algorithm}

The developed policy synthesis method is summarized in Alg.~\ref{algo:filter}. 
Essentially, the sets of applicable actions are iteratively shrunk according to the priority order of the goal sets. First, the actions that do not maximize the probability to reach the goal set $G_i$ are filtered (line~\ref{line:filter1}). Then, among the remaining actions, the actions that do not minimize the expected number of steps to reach $G_i$ are filtered (line~\ref{line:filter2}). After executing the main loop for each goal set, the policy that minimize the infinite horizon value function is computed (line~\ref{line:optimal}). % Next, the details of these steps are given. 

The iterative filtering process starts from $app$, i.e. $app_0 = app$ (see Sec.~\ref{sec:MDP}). Here, we first describe the computation of $app_i$ from $app_{i-1}$ with respect to the goal set $G_i$ and the corresponding optimization criteria $\oc{i}$. The maximal probability $P_{n, i}^{\star}(s)$ to reach $G_{i}$ w.r.to $app_{i-1}$ within $n$-steps is 
\begin{align}
	P_{n,i}^{\star}(s) & = \max_{a \in app_{i-1}(s)} \sum_{s' \in S} T(s, a, s') P_{n-1,i}^{\star}(s')  \label{eq:prgoGi} \\ 
	P_{0,i}^{\star}(s) &= 0 \text{ for } s \in S \setminus G_i, P_{0,i}^{\star}(s) = 1 \text{ for } s \in G_i. \nonumber
\end{align}
As shown by~\cite{s3p}, the sequence $P_{n, i}^{\star}$ converge to $P_{\infty, i}^{\star}$ that is the probability to reach $G_i$ under the optimal policy $\pi_{i'}^\star$ such that $\pi_{i'}^\star(s) \in app_{i-1}(s)$ for each $s\in S$ (see~\eqref{eq:fprgo}). Thus, we first compute the optimal probability $P_{\infty, i}^{\star}$ to reach $G_i$ from $app_{i-1}$ as in~\eqref{eq:prgoGi}, then filter the applicable actions that do not attain $P_{\infty, i}^{\star}$ (line~\ref{line:filter1} of Alg.~\ref{algo:filter}): 
\begin{align}
    app_{i'}(s) = & \{a \in app_{i-1}(s) \mid \label{eq:filter1} \\ 
    & P_{\infty, i}^{\star}(s) = \sum_{s' \in S}T(s,a,s')P_{\infty, i}^{\star}(s')\} \nonumber 
\end{align}
Note that if $P_{\infty, i}^{\star}(s) = 0$, then $app_{i'}(s) = app_{i-1}(s)$ via~\eqref{eq:filter1}. Next, we compute the optimal expected number of steps to reach $G_i$ over the paths that reach $G_i$ within $n$-steps using the filtered sets of applicable actions $app_{i'}$ that have the optimal goal-probability in $app_{i-1}$. In particular, we compute the optimal average cost to go function~\eqref{eq:Cgo} with respect to $P_{\infty, i}^{\star}$, $c_i$~\eqref{eq:Gi_cost} and $app_{i'}$:
\begin{align}
	\label{eq:Cgoi} 
	C_{n,i}^{\star}(s) =&
	\min_{a \in app_{i'}(s)}\frac{1}{ P_{\infty,i}^{\star}(s) } \times \\
	& \sum_{s' \in S} T(s, a, s') P_{\infty,i}^{\star}(s') (c_i(s,a,s') + C_{n-1,i}^{\star}(s')) \nonumber 
\end{align}
with $C_{0, i}^{\star}(s) = 0$, $\forall s \in S$. Since $c_i(s,a,s') = 1$ for each $S \setminus G_i$, $C_{n, i}^{\star}$ converges to $C_{\infty, i}^{\star}$ (Thm.3 from~\cite{s3p}). For a state $s\in S$, $C_{\infty, i}^{\star}(s)$  is the expected length of a path that originate from $s$ and end in $G_i$ under the optimal policy $\pi_i^\star$ w.r.to $app_i$, i.e, $\pi_i^\star(s) \in app_i(s)$ for each $s\in S$. Note that, the cost to go function~\eqref{eq:Cgoi} is only defined for the states $s\in S$ with a non-zero probability to reach $G_i$, i.e., when $P_{\infty,i}^{\star}(s) > 0$. After we compute the infinite horizon optimal cost $C_{\infty, i}^{\star}$, we filter the actions with non-optimal cost (line~\ref{line:filter2} of Alg.\ref{algo:filter}): 
\begin{align}
app_i(s) & = app_{i'}(s)  \quad \text{  if } P_\infty^{G_i}(s) = 0, \text{ otherwise } \label{eq:filter2} \\
app_i(s) & = \{ a \in app_{i'}(s) \mid C_{\infty, i}^{\star}(s) =  \frac{1}{ P_\infty^{G_i}(s) } \times \nonumber \\
& \sum_{s' \in S} T(s, a, s') P_{\infty}^{G_i}(s') (c_i(s,a,s') + C_{\infty,i}^{\star}(s')) \} \nonumber
\end{align}

We claim that any policy $\pi$ generated from $app_i$,  i.e., $\pi(s) \in app_{i}(s)$ for each $s \in S$, satisfies the optimization criteria $\oc{1}, \ldots, \oc{i}$ in this order and any optimal policy w.r.to $\oc{1}, \ldots, \oc{i}$ can be generated from $app_{i}$. Thus, the set of all optimal policies $\Pi_i$ as defined in~\eqref{eq:PolicySets} w.r.to $\oc{1}, \ldots, \oc{i}$ is:
\begin{equation}
    \label{eq:Pifromapp}
    \Pi_i = \{\pi \mid \pi(s) \in app_i(s) \}
\end{equation}

The claim that a policy $\pi$ generated from $app_i$ is optimal simply follows from the existence of an optimal stationary policy for the considered objectives~\citep{s3p}, and the iterative construction steps~\eqref{eq:filter1} and~\eqref{eq:filter2}, i.e., if $a \in app_j(s)$ then $a$ achieves the optimal goal-probability among $app_{j-1}(s)$ and the optimal expected number of steps to reach among $app_{j-1'}(s)$. 
The filtering steps guarantee that $app_j(s) \subseteq app_{i}(s)$ for each $j <i$ and $s \in S$. Thus, it holds that for any $\pi \in \Pi_i$ from~\eqref{eq:Pifromapp}, the probability to reach $G_j$ in $n$-steps under policy $\pi$, $P_{n,j}^\pi(s)$, equals to the maximal probability to reach $G_j$, i.e., $P_{n,j}^\star(s)$, among the policies that satisfy $\oc{1}, \ldots, \oc{j-1}$. The last argument follows by iterative applications starting from $i=1$ and the subset relation among $app(j)$ and $app(i)$. Furthermore, with a symmetric argument, we deduce that any $\pi \in \Pi_i$ from~\eqref{eq:Pifromapp} is optimal for each $G_j$, $j \leq i$, with respect to the expected number of steps to reach the goal sets.

The claim that each optimal policy satisfying $\oc{1}, \ldots, \oc{i}$ is included in $\Pi_i$ from~\eqref{eq:Pifromapp} can be seen by a contradiction argument. Assume that $\pi \not \in \Pi_i$ is an optimal stationary policy. Thus, $\pi(s) \not \in app_{i}(s)$ for some $s$. By the optimality of each $\pi' \in \Pi_i$~\eqref{eq:Pifromapp} with respect to $\oc{1}, \ldots, \oc{i}$, we reach that $\sum_{s' \in S} T(s,\pi(s),s')P_{\infty,j}^\star(s') = \sum_{s' \in S} T(s,\pi'(s),s')P_{\infty,j}^\star(s')$ for each $j \leq i$. Thus, $\pi(s)$ can not be filtered via~\eqref{eq:filter1}.
With a similar argument on~\eqref{eq:filter2}, we conclude that such a policy $\pi$ does not exists. 

The previous discussion yields that any policy generated from $app_n$ satisfies $\oc{1}, \ldots, \oc{n}$ in this order. Thus, as the final step, we synthesize the policy minimizing $V_\infty$~\eqref{eq:value} from $app_n$:

\begin{align}
  \pi^\star(s) &= \argmin_{a \in app_n(s)}
        \sum_{s' \in s} T(s, \pi(s), s')( c(s, a, s') + \gamma V_{\infty}^{\star}(s') ) \nonumber 
\end{align}
where 
\begin{align}
        V_n^{\star}(s) &=
        \min_{a \in app_n(s)}
        \sum_{s' \in s} T(s, \pi(s), s') ( c(s, a, s') + \gamma V_{n-1}^{\star}(s') ) \nonumber \\
        V_0^{\star}(s) &= 0 \nonumber 
\end{align}

%% file: Sections/mdp_construct.tex
% After an earthquake, a complete blackout is expected in the distribution system, i.e., all breakers are open to protect the system. During the restoration process, starting from the energy sources (transmission grid or distributed energy resources) energization actions are applied iteratively and the aim is to energize the whole system as soon as possible. Black-start is already a hard problem since electrical and topological constraints have to be considered. This problem becomes even more complex after an earthquake as the field instruments can be destructed. In~\cite{Gol2019}, the synthesis of an optimal restoration strategy problem is mapped to an MDP policy synthesis problem by using the probability of failure (i.e. destruction) values of the field instruments. These are computed by using the peak ground acceleration values recorded during the earth quake \red{earthquake?}.

The proposed goal oriented policy synthesis method is applied to an MDP that models restoration of an earthquake damaged distribution system~\citep{Gol2019}.  In the MDP model $M=(S,A,T,c)$, a state $s = (s^1, \ldots, s^N) \in S$ represents the health statuses of each system component (bus), and $s^i$ is the state of $i$-th bus in $s$. A bus can be in unknown ($U$), damaged ($D$) or energized ($E$) state. After the earthquake, all breakers are open and the statuses are unknown, thus the initial state is $s_1 = (U,\ldots,U)$. An action $a \subset \{1,\ldots, N\} \in A$ represents the set of busses that can be energized simultaneously and an energization action can only be applied to a bus that is in $U$ state. Furthermore, topological and electrical constraints limit the possible actions and they are integrated via $app: S \to 2^A$. The topological constraints include the connectivity to an energized bus or energy source, avoidance of generating an energized loop and preserving a minimum distance for the buses that can be energized simultaneously.
When an action $a \in A$ is applied in state $s=(s^{1},\ldots, s^{N})$, the MDP can transition to a state $s'=(s'^{1},\ldots, s'^{N})$ such that $s'^{i} \in \{E,D\}$ for each $i\in a$ and $s'^{i} = s^i$ for each $i \not \in a$, i.e., if a bus is tried to be energized, then its status can be $E$ or $D$ after the transition, and the statuses of the remaining buses do not change. The transition probabilities are computed with respect to the probability of failure values of the corresponding components. The cost is defined as the number of unenergized buses, i.e., $c(s,a,s')$ is the number of buses that are in $U$ or $D$ status in $s$.
Further details on the constraints and the model construction can be found in~\citep{Gol2019,Arpali2019}.

\cite{Gol2019} minimize the overall restoration time, whereas, \cite{Arpali2019} minimize the average energization time for each bus. 
In a post-disaster scenario, fast restoration of electricity for some buses, thus the corresponding buildings/infra-structure, can be more important than others. For example, energization of each hospital or energization of at least one base station serving an area with collapsed buildings can be prioritized.
Note that given a set of buses $B \subset \{1,\ldots,n\}$, in the first example we require each bus $i \in B$ to be energized (min-max case), whereas, in the second example we require at least one bus $i \in B$ to be energized (min-min case). 
Next, given a sequence of  prioritization sets $\textbf{B} = [B_1, \ldots, B_m]$ and their properties (min-min or min-max), we describe how we generate a sequence of goal sets $\textbf{G} = [G_1, \ldots, G_n]$ and apply the proposed goal-oriented synthesis method to $M$ and $\textbf{G}$.
% \red{did we specify $B_i \cap B_j = \emptyset, i \neq j$ anywhere? Is it important?}
% Why would it be? absorbing olmasi yeterli bizim icin, kesisimi olmasi siralamaya gore sonucu etkiler buyuk ihtimalle. Optimality acisindan onemli degil.  

\begin{remark}\label{remark:structure}
The construction steps ensure that the resulting MDP is acyclic and it includes states $s \in S$ for which no restoration action is possible. To avoid blocking states, a self transition with a special input is added to such states, i.e., $T(s, \emptyset, s) = 1$. Furthermore, due to the particular structure of the MDP, for any set of policies $\Pi$, the probability of to reach a goal set $G$ is the same for a state $s$. Thus, for any state $s\in S$ and policy $\pi \in \Pi$, the probability to reach $G$ under policy $\pi$, $P^{\pi}_{\infty}(s)$, equals to the maximal probability to reach $G$ over $\Pi$, i.e., $P^{\pi}_{\infty}(s) = \max_{\pi' \in \Pi} P^{\pi'}_{\infty}(s)$.
Due to this property of the MDP, the first filtering step (line~\ref{line:filter1} of Alg.~\ref{algo:filter}) is redundant and it is not applied in the policy synthesis.
\end{remark}

%% file: Sections/maxmin.tex
\subsection{Goal Sequence: Min-max case}
\label{sub:maxmin}

Given a set of buses $B \subset \{1,\ldots,N\}$, in the min-max case, the goal is to energize each bus $ i\in B$ within the minimum amount of time. Particularly, we aim at minimizing the maximum amount of time to energize a bus from $B$. Furthermore, it might not be possible to energize all the buses from $B$, i.e., some of them can be damaged or unreachable due to the other damaged buses. In such a case, it is desired to minimize the energization time for the remaining buses. To achieve this goal, we generate a sequence of goal sets $\textbf{G} = [G_1, \ldots, G_{|B|}]$ of length $|B|$ from the prioritization set $B$ such that $G_i \subset S$ is the set of MDP states in which at least $b_i = |B|-(i-1)$ buses from $B$ are energized: 
\[ G_i = \{ s \in S \mid \ \  \left |\{i \in B \mid s^i = E\} \right | \ \ \geq b_i \} \]
In particular, $G_1 = \{ s \in S \mid s^i = E \textrm{ for each } i \in B \}$ and $G_{|B|} = \{ s \in S \mid s^i = E \textrm{ for some } i \in B \}$. Note that each $G_i$ is absorbing since the status of a bus can not change to $D$ or $U$ from $E$. 

As highlighted in Remark~\ref{remark:structure}, each policy results in the same infinite horizon probability to reach a goal set. Thus, the optimal policy obtained from $\textbf{G}$ via Alg.~\ref{algo:filter} primarily minimizes the expected time to reach $G_1$, which in general, results in minimizing the expected energization time of the furthest bus from $B$. The use of the goal sequence instead of only $G_1$ has two major advantages. First, among the policies that are optimal w.r.to $G_1$, the policies that energize subsets of $B$ within the least expected number of steps are selected.  Second, in a state $s$, some of the busses $B' \subset B$ can be unreachable or damaged ($s^i = D$ for some $i \in B'$). In such a case, $P_{\infty,j}^\star(s)$~\eqref{eq:prgoGi} is $0$  for each $j < |B| - |B'|$. Since only the paths that lead to the given goal set with positive probability are considered at each stage of the filtering~\eqref{eq:filter2}, the actions applicable in $s$ are not filtered with respect to the goal sets $G_1, \ldots, G_{|B| - |B'|}$. 
However, the use of the goal sequence ensures that the actions applicable $s$ are filtered to minimize the expected energization time for the remaining buses, $B \setminus B'$. Thus the use of the goal sequence, and goal-probability in~\eqref{eq:filter2} allow us to minimize the expected energization time of all buses in $B$ in a best effort manner. Both advantages are illustrated in the example.

\subsection{Goal Sequence: Min-min Case}
\label{sub:minmin}

Given a set of buses $B \subset \{1,\ldots,N\}$, in the min-min case, the goal is to energize a bus $ i\in B$ within the minimum amount of time. To achieve this goal, we define a single goal set (i.e. a sequence  $\textbf{G} = [G]$) from $B$:
\[G = \{s\in S \mid s^i = E \textrm{ from some } i \in B \} \]
The optimal policy minimizes the expected time to energize a bus from $B$. As mentioned previously, this case can be used when different buses can serve for the same purpose, e.g., base stations covering the same area, redundant buses feeding the same building etc. 

Given a sequence of  prioritization sets $\textbf{B} = [B_1, \ldots, B_m]$ over the buses ($B_i \subset \{1,\ldots, N\}$) and their optimization  properties (max-min or min-max), we construct a goal sequence $\textbf{G}_i$ for each prioritization set $B_i$ as described in Sec.~\ref{sub:maxmin} and~\ref{sub:minmin} with respect to its optimization property, and then concatenate each $\textbf{G}_i$ in the given order to obtain the goal sequence $\textbf{G} = [\textbf{G}_1, \ldots, \textbf{G}_m]$ (observe that $|\textbf{G}| \geq m$). Finally, we generate the optimal strategy for $M$ from $\textbf{G}$ via Alg.~\ref{algo:filter}. Once the applicable actions are filtered w.r.to the goal sequence $\textbf{G}$, a policy minimizing the average expected restoration time is synthesized w.r.to the remaining sets of applicable actions.

%% file: Sections/example/main.tex
\subsection{Sample System}%
\label{sub:example}

In this section, the developed method is illustrated over a sample distribution system shown in Fig.~\ref{fig:example_system}. 
The system has $N=8$ buses and a single energy source. Only bus-1 is connected to the source. The MDP generated from this distribution system has $126$ states and $37$ of them are terminal states, i.e., states with self-loops (see Remark~\ref{remark:structure}). The initial state of this MDP is $s_1 = (U,U,U,U,U,U,U,U)$. 
% As it is not practical to demonstrate all $126$ states here, all of these states will not be explored during the solution.
% Instead we will focus on only a portion of the MDP, starting from the initial state $s_1$, in which all buses are in unknown status.

% Each node represents a bus. The index of each bus is written in the node, and probability of failure is below the node. There's only one energy source: $E_0$, and only component 1 is connected to it.

% Figure \ref{fig:example_system} depicts the distribution system, alongside with the probability of failure $P_f$ of each component. Component 1 is connected to the transmission grid, and none of them is energized.

\input{Sections/example/e0}
\input{Sections/example/s0}

\subsection{17-Bus Distribution System}

A real-life medium sized distribution system is shown in Fig.~\ref{fig:mid_size}. The system is connected to the transmission grid via buses 1 and 17. The MDP generated from this system has $9487$ states. For this example, we run synthesis algorithms by~\cite{Gol2019}
 and~\cite{Arpali2019}, and our algorithm for two prioritization sequences and report the expected cost values for the goal sets. 
 
 % we consider two prioritization sequences, $\textbf{B}_1$ and $\textbf{B}_2$, 
% From the initial state where every bus is in unknown state, the buses are energized starting from these buses in the opposite corners. The generated MDP corresponding to this system has 9487 states which include 2316 terminal states.

\input{Sections/example/17bus}

% On this example, the average length of a path from initial state to each $G_i$ for the optimal policies generated by different algorithms will be compared under different $\textbf{B}$ and goal sequence (min-max/min-min) configurations.

\begin{table}[H]
\centering
\caption{Comparison of Algorithms}
\begin{tabular}{l|p{0.2\linewidth} p{0.2\linewidth} p{0.2\linewidth}}
    %\multirow{2}{*}{ } & \multicolumn{3}{c}{$C_\infty^{G_i, \pi}$ induced by $c_i$} \\
                              & Our method & \cite{Arpali2019} & \cite{Gol2019} \\
                              \hline
%    $\textbf{G}_1$: $G_1$  &    6.394984 &    7.438871  &   7.583072 \\
% Hocam buraya $C_\infty^{G..}$ yazabiliriz? OK 
    $\textbf{G}_1$: $C_\infty^{G_1, \pi}$  &    6.3950 &    7.4389  &   7.5831 \\
%   $\textbf{G}_1$: $C_\infty^{G_2, \pi}$  &    6.600886 &    7.610802  &   7.274440 \\
    $\textbf{G}_1$: $C_\infty^{G_2, \pi}$  &    6.6009 &    7.6109  &   7.2744 \\
%    $V_{17}^\pi$           &  208.935342 &  208.499613  & 208.679696 \\
    $V_{17}^\pi$           &  208.94 &  208.50  & 208.68 \\
                              \hline
%    $\textbf{G}_2$: $G_1$  &    3.700849 &    4.574003  &    4.591947 \\
%    $\textbf{G}_2$: $G_2$  &    7.620298 &    7.438871  &    7.583072 \\
%    $\textbf{G}_2$: $G_3$  &    7.562127 &    7.610802  &    7.274440 \\
    $\textbf{G}_2$: $C_\infty^{G_1, \pi}$  &    3.7009 &    4.5740  &    4.5920 \\
    $\textbf{G}_2$: $C_\infty^{G_2, \pi}$  &    7.6203 &    7.4389  &    7.5831 \\
    $\textbf{G}_2$: $C_\infty^{G_3, \pi}$  &    7.5621 &    7.6108  &    7.2744 \\
%    $V_{17}^\pi$           &  209.745811 &  208.499613  & 208.679696 \\
    $V_{17}^\pi$           &  209.75 &  208.50  & 208.68 \\
  %   \hline
    
\end{tabular}
\label{compare:Bus17All}
\end{table}
% discount 1 horizon 17
% \green{ Ilker: Updated table \ref{compare:Bus17All}, fixed the errors in Arpali and Gol, added $V_n^\pi$ discount = 1, horizon = 17 }

First, we consider a singe priority set $\{ 6, 12 \}$ with min-max setting. The corresponding goal sequence is $\textbf{G}_1 = [ G_1, G_2 ]$ where $G_1$ is the set of states in which both bus-$6$ and bus-$12$ are energized, and $G_2$ is the set of states in which at least one of them is energized. The expected length of the paths that reach the goal sets for different policies are reported in the first two rows of Table~\ref{compare:Bus17All}. Even though $G_1 \subset G_2$, the expected time to reach $G_2$ is higher since the expected cost is computed over the paths that reach the goal set. 

Next, we add a new set $\{3,10\}$ with higher priority and min-min setting to the sequence, $\textbf{B} =[ \{3,10\}, \{6, 12\}]$. The corresponding goal sequence is $\textbf{G}_2 = [ G_1, G_2 , G_3]$, where $G_1$ is the set of states in which bus-$3$ or bus-$10$ is energized, and $G_2$ and $G_3$ are same as $G_1$ and $G_2$ from $\textbf{G}_1$, respectively. The results are shown in the second part of  Table~\ref{compare:Bus17All}. Note that 
as $\{3,10\}$ is prioritized over $\{6, 12\}$, the costs for the goal sets obtained from $\{6,12\}$ are increased compared to the first case.
As seen in the table, for each case the proposed method reduces the expected time to reach the prioritized set $G_1$ compared to~\cite{Gol2019} and~\cite{Arpali2019}. However, the expected time to reach other goal sets might increase (e.g. see $\textbf{G}_2$) since the expected time to reach $G_2$ is minimized only among the policies that minimize the expected time to reach $G_1$.

\cite{Arpali2019} minimize the finite horizon value function $V_{17}^\pi$ over $c$ from $M$. This value for our policy and the policies from~\cite{Gol2019} and~\cite{Arpali2019} are reported in Table~\ref{compare:Bus17All}. It is slightly increased as we prioritize the goal sets, and then find the optimal policy according to the same cost function.

% \red{it doesn't reduce for each case, sometimes it increases $G_2$ to reach $G_1$ faster}

% EBRU: Bu kisim repetition olmus. 
% \blue{ As seen in the table, for each case the proposed method reduces the expected time to reach $G_1$ set compared to ~\cite{Gol2019} and~\cite{Arpali2019}. However, the expected time to reach other goal sets might increase, as in the case of $\textbf{G}_2$. This happens because reaching $G_1$ as fast as possible is the most important optimization criterion. Since expected time to reach $G_2$ is minimized only among the policies that minimize expected time to reach $G_1$, expected time to reach $G_2$ can increase compared to a policy that minimizes the value function without any prioritization (as in ~\cite{Arpali2019}). }

\begin{table}[H]
\centering
\caption{Comparison of Algorithms for the System in Fig.~\ref{fig:mid_size}, for all priority sets $\{ i, j, k \}, i \neq j \neq k$, min-max}
\begin{tabular}{l l|p{0.2\linewidth} p{0.2\linewidth} p{0.2\linewidth}}
    & & \small{Our method} & \cite{Arpali2019} & \cite{Gol2019} \\
                              \hline
    \multirow{2}{*}{$C_\infty^{G_1, \pi}$} 
    & mean & 5.7745 & 6.5741 & 6.5404  \\
    & SD  & 0.9354 & 1.1431 & 1.1413  \\
%    & min  & 2.0      & 2.0      & 2.0        \\
%    & max  & 7.8383 & 8.3849 & 8.3222  \\
                              \hline
    \multirow{2}{*}{$C_\infty^{G_2, \pi}$} 
    & mean & 4.8137 & 5.0933 & 5.1136 \\
    & SD  & 1.3868 & 1.5421 & 1.5713 \\
%    & min  & 1.0589 & 1.0833 & 1.0677 \\
%    & max  & 7.2721 & 8.1034 & 7.9457 \\
                              \hline
    \multirow{2}{*}{$C_\infty^{G_3, \pi}$} 
    & mean & 2.7698 & 2.8061 & 2.8133 \\
    & SD  & 1.5548 & 1.5969 & 1.6107 \\
%    & min  & 1.0      & 1.0      & 1.0      \\
%    & max  & 6.8853 & 7.0598 & 7.2676 \\
                              \hline
    \multirow{2}{*}{$V_{17}^{\pi}$} 
    & mean & 209.01 & 208.50 & 208.68 \\
    & SD  &   0.307 & 0.0        & 0.0        \\
%    & min  & 208.50 & 208.50 & 208.68 \\
%    & max  & 210.76 & 208.50 & 208.68 \\
%    & mean & 209.005017 & 208.499613 & 208.679696 \\
%    & std  &   0.307007 & 0.0        & 0.0        \\
%    & min  & 208.502649 & 208.499613 & 208.679696 \\
%    & max  & 210.761389 & 208.499613 & 208.679696 \\
\end{tabular}
\label{compare:mid_size2}
\end{table}

To get a better insight into the behavior of our method and the previous methods, we test these algorithms on all priority sets with 3 buses with max-min setting, i.e., all sets $\{ i, j, k \}$ where $i, j, k \in \{1, \ldots, 17\}, i \neq j, j \neq k, i \neq k$ (680 cases), and report the results in Table \ref{compare:mid_size2}. Similar to the previous examples, all three buses are energized in $G_1$, at least two of them are energized in $G_2$ and at least one of them is energized in $G_3$. 
Since a considerable portion of these cases prioritize buses that are close to the transmission grid (e.g. buses 1,2, 13-15, 17), any policy must energize them to reach others. % Besides, the optimal policies from previous papers have a particular preference towards some buses which have a lower $V$ values or which are closer to a terminal state, and these buses are prioritized in some of these cases. 
Consequently, even though our policy results in better expected time to reach $G_1$ (also $G_2$, $G_3$), the average difference is not large (mean values). For the goal set with the highest priority ($G_1$), the maximum reduction of the cost ($C_\infty^{G_1, \pi}$) is $27\%$ compared to~\cite{Arpali2019} (our cost $5.7042$, their cost $7.8169$) that is observed for the bus set $\{2, 6, 16\}$. % The mean value for $V_{17}^\pi$ (the objective from~\cite{Arpali2019}) is also reported. 
Finally, as expected, introducing and prioritizing other optimization criteria effect $V_{17}^\pi$ (the objective from~\cite{Arpali2019}) negatively, however the average difference is quite small.

%% file: Sections/example/e0.tex
\begin{table}[ht]
  \begin{minipage}[b]{0.65\linewidth}
    \centering
\tikzset{every picture/.style={line width=0.75pt}} %set default line width to 0.75pt        

\begin{tikzpicture}[x=0.75pt,y=0.75pt,yscale=-0.8,xscale=0.8]
%uncomment if require: \path (0,485); %set diagram left start at 0, and has height of 485

%Straight Lines [id:da8618963250953409] 
\draw    (390.23,70.24) -- (390.07,119.64) ;
%Rounded Rect [id:dp6213010722349394] 
\draw  [fill={rgb, 255:red, 0; green, 0; blue, 0 }  ,fill opacity=1 ][line width=0.75]  (304.41,68.69) .. controls (304.41,68.12) and (304.87,67.66) .. (305.44,67.66) -- (475.01,67.66) .. controls (475.58,67.66) and (476.04,68.12) .. (476.04,68.69) -- (476.04,71.78) .. controls (476.04,72.35) and (475.58,72.82) .. (475.01,72.82) -- (305.44,72.82) .. controls (304.87,72.82) and (304.41,72.35) .. (304.41,71.78) -- cycle ;
%Shape: Square [id:dp5863785149490699] 
\draw  [fill={rgb, 255:red, 255; green, 255; blue, 255 }  ,fill opacity=1 ] (386.29,77.05) -- (393.95,77.05) -- (393.95,84.71) -- (386.29,84.71) -- cycle ;
%Shape: Square [id:dp045091070594560634] 
\draw  [fill={rgb, 255:red, 255; green, 255; blue, 255 }  ,fill opacity=1 ] (386.38,106.29) -- (394.05,106.29) -- (394.05,113.95) -- (386.38,113.95) -- cycle ;
%Rounded Rect [id:dp615002391369306] 
\draw  [fill={rgb, 255:red, 0; green, 0; blue, 0 }  ,fill opacity=1 ] (360.89,118.22) .. controls (360.89,117.7) and (361.31,117.28) .. (361.83,117.28) -- (418.3,117.28) .. controls (418.82,117.28) and (419.24,117.7) .. (419.24,118.22) -- (419.24,121.05) .. controls (419.24,121.57) and (418.82,121.99) .. (418.3,121.99) -- (361.83,121.99) .. controls (361.31,121.99) and (360.89,121.57) .. (360.89,121.05) -- cycle ;
%Straight Lines [id:da9291343838891761] 
\draw    (390.18,54.6) -- (390.23,70.24) ;
%Straight Lines [id:da08932567037190531] 
\draw    (309.23,71.24) -- (309.07,120.64) ;
%Shape: Square [id:dp22421329565504866] 
\draw  [fill={rgb, 255:red, 255; green, 255; blue, 255 }  ,fill opacity=1 ] (305.29,78.05) -- (312.95,78.05) -- (312.95,85.71) -- (305.29,85.71) -- cycle ;
%Shape: Square [id:dp5448433291740307] 
\draw  [fill={rgb, 255:red, 255; green, 255; blue, 255 }  ,fill opacity=1 ] (305.38,107.29) -- (313.05,107.29) -- (313.05,114.95) -- (305.38,114.95) -- cycle ;
%Rounded Rect [id:dp8096604447519524] 
\draw  [fill={rgb, 255:red, 0; green, 0; blue, 0 }  ,fill opacity=1 ] (279.89,119.22) .. controls (279.89,118.7) and (280.31,118.28) .. (280.83,118.28) -- (337.3,118.28) .. controls (337.82,118.28) and (338.24,118.7) .. (338.24,119.22) -- (338.24,122.05) .. controls (338.24,122.57) and (337.82,122.99) .. (337.3,122.99) -- (280.83,122.99) .. controls (280.31,122.99) and (279.89,122.57) .. (279.89,122.05) -- cycle ;
%Straight Lines [id:da8539599913306865] 
\draw    (472.23,70.24) -- (472.07,119.64) ;
%Shape: Square [id:dp8588725074642922] 
\draw  [fill={rgb, 255:red, 255; green, 255; blue, 255 }  ,fill opacity=1 ] (468.29,77.05) -- (475.95,77.05) -- (475.95,84.71) -- (468.29,84.71) -- cycle ;
%Shape: Square [id:dp6397950747974741] 
\draw  [fill={rgb, 255:red, 255; green, 255; blue, 255 }  ,fill opacity=1 ] (468.38,106.29) -- (476.05,106.29) -- (476.05,113.95) -- (468.38,113.95) -- cycle ;
%Rounded Rect [id:dp2420267575739321] 
\draw  [fill={rgb, 255:red, 0; green, 0; blue, 0 }  ,fill opacity=1 ] (442.89,118.22) .. controls (442.89,117.7) and (443.31,117.28) .. (443.83,117.28) -- (500.3,117.28) .. controls (500.82,117.28) and (501.24,117.7) .. (501.24,118.22) -- (501.24,121.05) .. controls (501.24,121.57) and (500.82,121.99) .. (500.3,121.99) -- (443.83,121.99) .. controls (443.31,121.99) and (442.89,121.57) .. (442.89,121.05) -- cycle ;
%Straight Lines [id:da25945457888561585] 
\draw    (390.23,118.24) -- (390.07,167.64) ;
%Shape: Square [id:dp5460146293280694] 
\draw  [fill={rgb, 255:red, 255; green, 255; blue, 255 }  ,fill opacity=1 ] (386.29,125.05) -- (393.95,125.05) -- (393.95,132.71) -- (386.29,132.71) -- cycle ;
%Shape: Square [id:dp5495758836375766] 
\draw  [fill={rgb, 255:red, 255; green, 255; blue, 255 }  ,fill opacity=1 ] (386.38,154.29) -- (394.05,154.29) -- (394.05,161.95) -- (386.38,161.95) -- cycle ;
%Rounded Rect [id:dp44014233113618606] 
\draw  [fill={rgb, 255:red, 0; green, 0; blue, 0 }  ,fill opacity=1 ] (360.89,166.22) .. controls (360.89,165.7) and (361.31,165.28) .. (361.83,165.28) -- (418.3,165.28) .. controls (418.82,165.28) and (419.24,165.7) .. (419.24,166.22) -- (419.24,169.05) .. controls (419.24,169.57) and (418.82,169.99) .. (418.3,169.99) -- (361.83,169.99) .. controls (361.31,169.99) and (360.89,169.57) .. (360.89,169.05) -- cycle ;
%Straight Lines [id:da6129481665808583] 
\draw    (309.23,119.24) -- (309.07,168.64) ;
%Shape: Square [id:dp6554892952119263] 
\draw  [fill={rgb, 255:red, 255; green, 255; blue, 255 }  ,fill opacity=1 ] (305.29,126.05) -- (312.95,126.05) -- (312.95,133.71) -- (305.29,133.71) -- cycle ;
%Shape: Square [id:dp9314669483740932] 
\draw  [fill={rgb, 255:red, 255; green, 255; blue, 255 }  ,fill opacity=1 ] (305.38,155.29) -- (313.05,155.29) -- (313.05,162.95) -- (305.38,162.95) -- cycle ;
%Rounded Rect [id:dp5377463262969964] 
\draw  [fill={rgb, 255:red, 0; green, 0; blue, 0 }  ,fill opacity=1 ] (279.89,167.22) .. controls (279.89,166.7) and (280.31,166.28) .. (280.83,166.28) -- (337.3,166.28) .. controls (337.82,166.28) and (338.24,166.7) .. (338.24,167.22) -- (338.24,170.05) .. controls (338.24,170.57) and (337.82,170.99) .. (337.3,170.99) -- (280.83,170.99) .. controls (280.31,170.99) and (279.89,170.57) .. (279.89,170.05) -- cycle ;
%Straight Lines [id:da7009542593382363] 
\draw    (472.23,118.24) -- (472.07,167.64) ;
%Shape: Square [id:dp05659107128897278] 
\draw  [fill={rgb, 255:red, 255; green, 255; blue, 255 }  ,fill opacity=1 ] (468.29,125.05) -- (475.95,125.05) -- (475.95,132.71) -- (468.29,132.71) -- cycle ;
%Shape: Square [id:dp8468473835881949] 
\draw  [fill={rgb, 255:red, 255; green, 255; blue, 255 }  ,fill opacity=1 ] (468.38,154.29) -- (476.05,154.29) -- (476.05,161.95) -- (468.38,161.95) -- cycle ;
%Rounded Rect [id:dp0974615153155628] 
\draw  [fill={rgb, 255:red, 0; green, 0; blue, 0 }  ,fill opacity=1 ] (442.89,166.22) .. controls (442.89,165.7) and (443.31,165.28) .. (443.83,165.28) -- (500.3,165.28) .. controls (500.82,165.28) and (501.24,165.7) .. (501.24,166.22) -- (501.24,169.05) .. controls (501.24,169.57) and (500.82,169.99) .. (500.3,169.99) -- (443.83,169.99) .. controls (443.31,169.99) and (442.89,169.57) .. (442.89,169.05) -- cycle ;
%Straight Lines [id:da2990101079868619] 
\draw    (390.23,167.24) -- (390.07,216.64) ;
%Shape: Square [id:dp9867162775581106] 
\draw  [fill={rgb, 255:red, 255; green, 255; blue, 255 }  ,fill opacity=1 ] (386.29,174.05) -- (393.95,174.05) -- (393.95,181.71) -- (386.29,181.71) -- cycle ;
%Shape: Square [id:dp13637410880340284] 
\draw  [fill={rgb, 255:red, 255; green, 255; blue, 255 }  ,fill opacity=1 ] (386.38,203.29) -- (394.05,203.29) -- (394.05,210.95) -- (386.38,210.95) -- cycle ;
%Rounded Rect [id:dp14300434792940875] 
\draw  [fill={rgb, 255:red, 0; green, 0; blue, 0 }  ,fill opacity=1 ] (360.89,215.22) .. controls (360.89,214.7) and (361.31,214.28) .. (361.83,214.28) -- (418.3,214.28) .. controls (418.82,214.28) and (419.24,214.7) .. (419.24,215.22) -- (419.24,218.05) .. controls (419.24,218.57) and (418.82,218.99) .. (418.3,218.99) -- (361.83,218.99) .. controls (361.31,218.99) and (360.89,218.57) .. (360.89,218.05) -- cycle ;
%Shape: Rectangle [id:dp5895447842365609] 
\draw   (372,32.67) -- (409.33,32.67) -- (409.33,54) -- (372,54) -- cycle ;
%Flowchart: Decision [id:dp895726539320798] 
\draw   (390.67,32.67) -- (409.33,43.33) -- (390.67,54) -- (372,43.33) -- cycle ;
%Straight Lines [id:da3515276970872637] 
\draw    (372,32.67) -- (409.33,54) ;
%Straight Lines [id:da12342902044929394] 
\draw    (409.33,32.67) -- (372,54) ;

% Text Node
\draw (313.13,60.09) node   [align=left] {1};
% Text Node
\draw (287.63,110.2) node   [align=left] {2};
% Text Node
\draw (367.67,109.67) node   [align=left] {4};
% Text Node
\draw (287,159.33) node   [align=left] {3};
% Text Node
\draw (449.67,158.67) node   [align=left] {8};
% Text Node
\draw (367.47,158.6) node   [align=left] {5};
% Text Node
\draw (367.8,206.93) node   [align=left] {6};
% Text Node
\draw (449.8,110.6) node   [align=left] {7};

\end{tikzpicture}

  \end{minipage}
  %\hfill
  \begin{varwidth}[b]{0.25\linewidth}
  {
      \renewcommand{\arraystretch}{1.2}
      \centering
      \begin{tabular}{c|c}
          Bus id & $P_f$ \\
          \hline
            1 & 0.125 \\
            2 & 0.500 \\ 
            3 & 0.250 \\ 
            4 & 0.500 \\ 
            5 & 0.500 \\ 
            6 & 0.500 \\ 
            7 & 0.125 \\ 
            8 & 0.125 \\ 
            \multicolumn{1}{c}{} \\
      \end{tabular}
    %\color{white}
    %\caption{
    %Probability of Failure
    %}
\hspace{0pt}
\vfill
  }
  \end{varwidth}%
  \vspace{8pt}

    \captionof{figure}{A distribution system and the probability of failure ($P_f$) values for each bus.}
    \label{fig:example_system}
\end{table}

%% file: Sections/example/s0.tex
%\subsubsection{Solution}%
% \label{sub:solution}

We consider a single priority set $B = \{3, 6\}$ and min-max optimization property (see Sec.~\ref{sub:maxmin}). Thus, we generate the goal sequence $\textbf{G} = [ G_1, G_2 ]$ where $G_1$ is the set of states in which both bus-$3$ and bus-$6$ are energized, and $G_2$ is the set of states in which bus-$3$ or bus-$6$ is energized. We run Alg.~\ref{algo:filter} on $M$ and $\textbf{G}$ and generate the optimal policy $\pi^\star$. Next, we illustrate this policy and the filtering steps over some states.

% Suppose that we have only one priority class, and in that class we have 2 components: 3 and 6. In this solution, we will be creating a policy that solves the max/min case. Hence, $\textbf{G} = [ G_1, G_2 ]$ where $G_1$ is the set of states in which both components 3 and 6 are energized, and $G_2$ is the set of states in which component 3 or component 6 is energized.

\begin{table}[H]
    
\caption{\\ Transitions from $s_1 = (U,U,U,U,U,U,U,U)$}
\begin{center}

\begin{tabular}{c|c|c}
    Action & Probability & Next State \\
    \hline
    \multirow{2}{*}{$\{1\}$} & 0.875 & $s_2 = (E,U,U,U,U,U,U,U)$ \\
                           & 0.125 & $s_3 = (D,U,U,U,U,U,U,U)$
\end{tabular}

\label{transitions:s1}
\end{center}
\end{table}

\begin{table}[H]
	\centering
	\caption{Optimal values for $s_1$}
	\begin{tabular}{ l|l l l l l }
		Action & $P_{\infty, 1}^{\star}$ & $P_{\infty,2}^{\star}$ & $C_{\infty, 1}^{\star}$ & $C_{\infty, 2}^{\star}$\\
		\hline
$\{1\}$ & $0.041016$ & $0.396484$ & $4.000$ & $4.000$ \\
\end{tabular}
	\label{distances:s1}
\end{table}

The transitions leaving the initial state $s_1$ and the optimal values for $s_1$ are shown in Tables~\ref{transitions:s1} and~\ref{distances:s1}, respectively. Since only bus-1 is connected to an energy source, there is only one applicable action $app(s_1) = \{ \{1\} \}$. Thus, $\pi^\star(s_1) = \{1\}$. The states reachable from $s_1$ are $s_2$ and $s_3$. $s_3$ is a terminal state with $app(s_3) = \{ \emptyset \}$ (see Remark~\ref{remark:structure}). The transitions and the optimal values for $s_2$ (for each action) are shown in Tables~\ref{transitions:s2} and~\ref{distances:s2}. As bus $1$ is connected to buses $\{2,4,7\}$ there are $3$ actions in $app(s_2)$.
  
% Table shows the transitions leaving $s_1$.5 Table  shows expected time to reach goals $G_1$ and $G_2$ (columns $C_{\infty, [1]}^{\star}$ and $C_{\infty, [2]}^{\star}$) from $s_1$ and the value function $V$ for each action.

% In the initial state, we only have one action to take, since there's only one component with energy source connected to it. $s_3$ is a dead-end state. Otherwise, if we end up in $s_2$ after taking this action, we would have different actions to choose from. As seen in Tables \ref{transitions:s2} and \ref{distances:s2}, we have 3 actions to choose from, since component 1 is connected to 3 different components in state $s_2$.

\begin{table}[H]
	\centering
	\caption{\\ Transitions from $s_2 = (E,U,U,U,U,U,U,U)$}
	\begin{tabular} {c|c|c}
		Action & Probability & Next State \\
		\hline
		\multirow{2}{*}{$\{4\}$} & 0.500 & $s_4 = (E,U,U,E,U,U,U,U)$ \\
                               & 0.500 & $s_{5} = (E,U,U,D,U,U,U,U)$ \\
                               \hline
        \multirow{2}{*}{$\{2\}$} & 0.500 & $s_{6} = (E,E,U,U,U,U,U,U)$ \\
                               & 0.500 & $s_{7} = (E,D,U,U,U,U,U,U)$ \\
                               \hline
		\multirow{2}{*}{$\{7\}$} & 0.875 & $s_{8} = (E,U,U,U,U,U,E,U)$ \\
                               & 0.125 & $s_{9} = (E,U,U,U,U,U,D,U)$ \\
	\end{tabular}
	\label{transitions:s2}
\end{table}

\begin{table}[H]
	\centering
	\caption{Optimal values for $s_2$}
	\begin{tabular}{ l|l l l l l }
		Action & $P_{\infty, 1}^{\star}$ & $P_{\infty, 2}^{\star}$ & $C_{\infty, 1}^{\star}$ & $C_{\infty, 2}^{\star}$\\
		\hline
		$\{4\}$ & $0.046875$ & $0.453125$  & $3.000$ & $3.000$  \\
		$\{2\}$ & $0.046875$ & -          & $4.000$  & -  \\
		$\{7\}$ & $0.046875$ & -          & $4.000$  & - 
\end{tabular}
\label{distances:s2}
\end{table}

We first filter the set of applicable actions with respect to $G_1$~\eqref{eq:filter2}, and reach $app_1(s_2) = \{\{4\}\}$ since $\{4\}$ is the only action attaining the optimal $C^\star_{\infty,1}$ value at $s_2$. Thus $\pi^\star(s_2) = \{4\}$. Note that $P_{\infty, 2}^\star(s_2)$ and  $C_{\infty,2}^\star(s_2)$ are not computed for actions $\{2\}$ and $\{7\}$ as they are filtered in the first iteration. 

% As we are trying to solve the max-min case, we first need to maximize the probability to energize both components, i.e., maximize $P_{\infty, [1]}^\star$, and then minimize the expected time to energize both components, i.e., minimize $C_{\infty, [1]}^{\star}$. All actions have the same $P_{\infty, [1]}^\star$ value, but only $a_1$ has the minimum $C_{\infty, [1]}^\star$. Therefore, we can safely say that the optimal policy for the max-min case should select $a_1$. However, if there were multiple actions with minimum $C_{\infty, [1]}^{\star}$, we would need to select an action which maximizes $P_{\infty, [2]}^\star$, minimizes $C_{\infty, [2]}^{\star}$, and minimizes $V$ respectively among these actions.

% Also observe that none of $P_{\infty, [2]}^\star$, $C_{\infty, [2]}^\star$ and $V$ are available for $a_2$ and $a_3$. This is because these actions were filtered in the first stage, and were never considered while calculating these functions.

\begin{table}[H]
	\centering
	\caption{\\ Transitions from $s_4 = (E,U,U,E,U,U,U,U)$}
	\begin{tabular} {c|c|c}
		Action & Probability & Next State \\
		\hline
		\multirow{4}{*}{$\{2,5\}$} & 0.250 & $s_{10} = (E,E,U,E,E,U,U,U)$ \\
& 0.250 & $s_{11} = (E,E,U,E,D,U,U,U)$ \\
& 0.250 & $s_{12} = (E,D,U,E,E,U,U,U)$ \\
& 0.250 & $s_{13} = (E,D,U,E,D,U,U,U)$ \\
\hline
\multirow{4}{*}{$\{5,7\}$} & 0.4375 & $s_{14} = (E,U,U,E,E,U,E,U)$ \\
& 0.4375 & $s_{15} = (E,U,U,E,D,U,E,U)$ \\
& 0.0625 & $s_{16} = (E,U,U,E,E,U,D,U)$ \\
& 0.0625 & $s_{17} = (E,U,U,E,D,U,D,U)$ \\
	\end{tabular}
	\label{transitions:s4}
\end{table}

\begin{table}[H]
	\centering
	\caption{Optimal values for $s_4$}
	\begin{tabular}{ l|l l l l l }
		Action & $P_{\infty, 1}^{\star}$ & $P_{\infty, 2}^{\star}$ & $C_{\infty, 1}^{\star}$ & $C_{\infty, 2}^{\star}$\\
		\hline
		$\{2,5\}$ & $0.09375$ & $0.531250$  & $2.000$  & $2.000$ \\
		$\{5,7\}$ & $0.09375$ & -         & $3.000$  & -\\
\end{tabular}
\label{distances:s4}
\end{table}

Next, we consider $s_4$. There are two actions in $app(s_4)$, each action tries to energize two buses simultaneously (subsets of these actions are omitted). The corresponding transitions and optimal values are shown in Tables~\ref{transitions:s4} and~\ref{distances:s4}, respectively. As $C_{\infty, 1}^{\star}$ is lower for $\{2,5\}$, $\{2,7\}$ is filtered in the first iteration and $\pi^\star(s_4) = \{2,5\}$. 

% Next, we will inspect the state $s_4$. As seen in Tables \ref{transitions:s4} and \ref{distances:s4}, we have 2 actions. $G_1$ is the goal set with the utmost priority. Although both actions have the same $P_{\infty, 1}^\star$, first filtering step will filter out $a_2$ because $a_1$ minimizes $C_{\infty, [1]}^{\star}$.

% Ebru: no need for the following case.
% After that, we only have one action in the state $s_{10} = (E,E,U,E,E,U,U,U)$. Component 1 attempts to energize component 7; component 2, component 3; and component 5, component 6. Since these components are not connected to any other unenergized component, this is our only action.

%Note that both $C_{\infty, [1]}^{\star}$ and $C_{\infty, [2]}^{\star}$ are equal to $1$,
%because if this action succeeds (leading to state $s_{43}$), all prioritized components will be energized in 1 step.

%For the sake of completeness, Tables \ref{transitions:s43} and \ref{distances:s43} describe the actions and their goal-cost and value functions from $s_{43}$.
%As we have energized the prioritized components already, both $C_{\infty, [1]}^{\star}$ and $C_{\infty, [2]}^{\star}$ are 0. After this point, we only minimize $V$, our last optimization criterion.

% \subsubsection{Less Than Ideal Cases}%
% Previously, we have inspected the case when all attempts to energize are successful. Next, we will examine the cases when the attempt to energize a component is unsuccessful.

Finally, we consider $s_{5}$ (see Table~\ref{transitions:s2}). Note that bus-$4$ is known to be damaged in $s_5$. Thus, it is not possible to energize the prioritized bus $6 \in B$. As a result, $P_{\infty,1}^{\star}(s_{5}) = 0$ and the filter w.r.to $G_1$ is not applied to $s_{5}$~\eqref{eq:filter2} (see Tables~\ref{transitions:s5} and~\ref{distances:s5}). Two actions are available in $s_{5}$, $app(s_{5}) = \{\{2\}, \{7\}\}$. Among these, only $\{2\}$ minimizes $C_{\infty, 2}^{\star}$ and $\{7\}$ is filtered ($\pi^{\star}(s_5) = \{2\}$). Thus, even though it is not possible to energize all the buses, the developed method minimizes the energization time for the remaining prioritized buses. Since the generated MDP has too many states to be illustrated, further details are not given.

 \begin{table}[H]
	\centering
	\caption{\\ Transitions from $s_5 = (E,U,U,D,U,U,U,U)$}
	\begin{tabular} {c|c|c}
		Action & Probability & Next State \\
		\hline
		\multirow{2}{*}{\{2\}} & 0.500 & $s_{18} = (E,E,U,D,U,U,U,U)$ \\
& 0.500 & $s_{19} = (E,D,U,D,U,U,U,U)$ \\
\hline
\multirow{2}{*}{$\{7\}$} & 0.875 & $s_{20} = (E,U,U,D,U,U,E,U)$ \\
& 0.125 & $s_{21} = (E,U,U,D,U,U,D,U)$ \\
	\end{tabular}
	\label{transitions:s5}
 \end{table}

 \begin{table}[H]
	\centering
		\caption{Optimal values for $s_{5}$}
	\begin{tabular}{ l|l l l l l }
		Action & $P_{\infty, [1]}^{\star}$ & $P_{\infty, [2]}^{\star}$ & $C_{\infty, [1]}^{\star}$ & $C_{\infty, [2]}^{\star}$\\
		\hline
		$\{2\}$ & $0$ & $0.375$ & N/A & $2.000$ \\
		$\{7\}$ & $0$ & $0.375$ & N/A & $3.000$        \\
\end{tabular}
\label{distances:s5}
\end{table}

%% file: Sections/example/17bus.tex
\begin{table}[ht]
  \begin{minipage}[b]{0.65\linewidth}
    \centering
\tikzset{every picture/.style={line width=0.75}} %set default line width to 0.75pt        

\begin{tikzpicture}[x=0.75pt,y=0.75pt,yscale=-0.75,xscale=0.8]
%uncomment if require: \path (0,489); %set diagram left start at 0, and has height of 489
%Shape: Rectangle [id:dp7490862728828036] 

%Rounded Rect [id:dp2426241565626941] 
\draw  [fill={rgb, 255:red, 0; green, 0; blue, 0 }  ,fill opacity=1 ][line width=0.75]  (369.73,39.59) .. controls (369.73,39) and (370.21,38.53) .. (370.8,38.53) -- (425.69,38.53) .. controls (426.28,38.53) and (426.75,39) .. (426.75,39.59) -- (426.75,42.77) .. controls (426.75,43.35) and (426.28,43.83) .. (425.69,43.83) -- (370.8,43.83) .. controls (370.21,43.83) and (369.73,43.35) .. (369.73,42.77) -- cycle ;
%Straight Lines [id:da5993624015309964] 
\draw    (398.24,41.18) -- (398.09,88.79) ;
%Rounded Rect [id:dp293025361022629] 
\draw  [fill={rgb, 255:red, 0; green, 0; blue, 0 }  ,fill opacity=1 ][line width=0.75]  (339.89,88.32) .. controls (339.89,87.79) and (340.32,87.37) .. (340.84,87.37) -- (453.43,87.37) .. controls (453.95,87.37) and (454.38,87.79) .. (454.38,88.32) -- (454.38,91.16) .. controls (454.38,91.68) and (453.95,92.11) .. (453.43,92.11) -- (340.84,92.11) .. controls (340.32,92.11) and (339.89,91.68) .. (339.89,91.16) -- cycle ;
%Shape: Rectangle [id:dp7246313378653004] 
\draw  [fill={rgb, 255:red, 255; green, 255; blue, 255 }  ,fill opacity=1 ] (394.42,47.74) -- (401.86,47.74) -- (401.86,55.13) -- (394.42,55.13) -- cycle ;
%Shape: Rectangle [id:dp6296074484082603] 
\draw  [fill={rgb, 255:red, 255; green, 255; blue, 255 }  ,fill opacity=1 ] (394.52,75.92) -- (401.95,75.92) -- (401.95,83.31) -- (394.52,83.31) -- cycle ;
%Rounded Rect [id:dp8860324618586426] 
\draw  [fill={rgb, 255:red, 0; green, 0; blue, 0 }  ,fill opacity=1 ] (330.36,137.6) .. controls (330.36,137.11) and (330.76,136.7) .. (331.25,136.7) -- (406.5,136.7) .. controls (407,136.7) and (407.4,137.11) .. (407.4,137.6) -- (407.4,140.29) .. controls (407.4,140.79) and (407,141.19) .. (406.5,141.19) -- (331.25,141.19) .. controls (330.76,141.19) and (330.36,140.79) .. (330.36,140.29) -- cycle ;
%Rounded Rect [id:dp27051227409909706] 
\draw  [fill={rgb, 255:red, 0; green, 0; blue, 0 }  ,fill opacity=1 ] (426.24,137.03) .. controls (426.24,136.52) and (426.65,136.12) .. (427.15,136.12) -- (481.92,136.12) .. controls (482.43,136.12) and (482.83,136.52) .. (482.83,137.03) -- (482.83,139.75) .. controls (482.83,140.25) and (482.43,140.66) .. (481.92,140.66) -- (427.15,140.66) .. controls (426.65,140.66) and (426.24,140.25) .. (426.24,139.75) -- cycle ;
%Straight Lines [id:da5721527896412115] 
\draw [color={rgb, 255:red, 0; green, 0; blue, 0 }  ,draw opacity=1 ]   (358.75,89.91) -- (358.38,137.43) ;
%Straight Lines [id:da7123453435287654] 
\draw    (454.38,91.16) -- (454.54,138.39) ;
%Rounded Rect [id:dp5994951863758078] 
\draw  [fill={rgb, 255:red, 0; green, 0; blue, 0 }  ,fill opacity=1 ] (285.7,186.29) .. controls (285.7,185.82) and (286.08,185.44) .. (286.55,185.44) -- (359.77,185.44) .. controls (360.24,185.44) and (360.62,185.82) .. (360.62,186.29) -- (360.62,188.85) .. controls (360.62,189.32) and (360.24,189.7) .. (359.77,189.7) -- (286.55,189.7) .. controls (286.08,189.7) and (285.7,189.32) .. (285.7,188.85) -- cycle ;
%Straight Lines [id:da5558077055825197] 
\draw [color={rgb, 255:red, 0; green, 0; blue, 0 }  ,draw opacity=1 ]   (406.5,141.19) -- (406.66,188.41) ;
%Rounded Rect [id:dp3670915876024279] 
\draw  [fill={rgb, 255:red, 0; green, 0; blue, 0 }  ,fill opacity=1 ] (389.51,186.12) .. controls (389.51,185.67) and (389.87,185.31) .. (390.32,185.31) -- (564.97,185.31) .. controls (565.42,185.31) and (565.79,185.67) .. (565.79,186.12) -- (565.79,188.56) .. controls (565.79,189.01) and (565.42,189.38) .. (564.97,189.38) -- (390.32,189.38) .. controls (389.87,189.38) and (389.51,189.01) .. (389.51,188.56) -- cycle ;
%Straight Lines [id:da26707234511587763] 
\draw    (332,140.87) -- (332.17,188.09) ;
%Shape: Rectangle [id:dp557282175930544] 
\draw  [fill={rgb, 255:red, 255; green, 255; blue, 255 }  ,fill opacity=1 ] (355.17,95.2) -- (362.61,95.2) -- (362.61,102.59) -- (355.17,102.59) -- cycle ;
%Shape: Rectangle [id:dp8218113681467967] 
\draw  [fill={rgb, 255:red, 255; green, 255; blue, 255 }  ,fill opacity=1 ] (354.9,125.76) -- (362.33,125.76) -- (362.33,133.15) -- (354.9,133.15) -- cycle ;
%Shape: Rectangle [id:dp015476401883822266] 
\draw  [fill={rgb, 255:red, 255; green, 255; blue, 255 }  ,fill opacity=1 ] (450.76,96.3) -- (458.19,96.3) -- (458.19,103.69) -- (450.76,103.69) -- cycle ;
%Shape: Rectangle [id:dp3323975538336237] 
\draw  [fill={rgb, 255:red, 255; green, 255; blue, 255 }  ,fill opacity=1 ] (450.76,125.49) -- (458.19,125.49) -- (458.19,132.88) -- (450.76,132.88) -- cycle ;
%Shape: Rectangle [id:dp4741904377669688] 
\draw  [fill={rgb, 255:red, 255; green, 255; blue, 255 }  ,fill opacity=1 ] (328.58,145.32) -- (336.01,145.32) -- (336.01,152.71) -- (328.58,152.71) -- cycle ;
%Shape: Rectangle [id:dp35450514122487653] 
\draw  [fill={rgb, 255:red, 255; green, 255; blue, 255 }  ,fill opacity=1 ] (328.58,174.78) -- (336.01,174.78) -- (336.01,182.17) -- (328.58,182.17) -- cycle ;
%Shape: Rectangle [id:dp7903714392899321] 
\draw  [fill={rgb, 255:red, 255; green, 255; blue, 255 }  ,fill opacity=1 ] (402.83,145.87) -- (410.26,145.87) -- (410.26,153.26) -- (402.83,153.26) -- cycle ;
%Shape: Rectangle [id:dp7310091053085304] 
\draw  [fill={rgb, 255:red, 255; green, 255; blue, 255 }  ,fill opacity=1 ] (402.83,175.06) -- (410.26,175.06) -- (410.26,182.45) -- (402.83,182.45) -- cycle ;
%Rounded Rect [id:dp6581642238527292] 
\draw  [fill={rgb, 255:red, 0; green, 0; blue, 0 }  ,fill opacity=1 ] (272,231.31) .. controls (272,230.81) and (272.41,230.4) .. (272.91,230.4) -- (327.68,230.4) .. controls (328.18,230.4) and (328.59,230.81) .. (328.59,231.31) -- (328.59,234.03) .. controls (328.59,234.53) and (328.18,234.94) .. (327.68,234.94) -- (272.91,234.94) .. controls (272.41,234.94) and (272,234.53) .. (272,234.03) -- cycle ;
%Straight Lines [id:da9467733601961377] 
\draw    (300.13,185.44) -- (300.29,232.67) ;
%Shape: Rectangle [id:dp43941154087918033] 
\draw  [fill={rgb, 255:red, 255; green, 255; blue, 255 }  ,fill opacity=1 ] (296.58,193.56) -- (304.01,193.56) -- (304.01,200.94) -- (296.58,200.94) -- cycle ;
%Shape: Rectangle [id:dp07207565464167498] 
\draw  [fill={rgb, 255:red, 255; green, 255; blue, 255 }  ,fill opacity=1 ] (296.25,219.85) -- (303.69,219.85) -- (303.69,227.24) -- (296.25,227.24) -- cycle ;
%Rounded Rect [id:dp5786688418424206] 
\draw  [fill={rgb, 255:red, 0; green, 0; blue, 0 }  ,fill opacity=1 ] (344.79,231.48) .. controls (344.79,230.98) and (345.19,230.57) .. (345.7,230.57) -- (400.47,230.57) .. controls (400.97,230.57) and (401.38,230.98) .. (401.38,231.48) -- (401.38,234.2) .. controls (401.38,234.71) and (400.97,235.11) .. (400.47,235.11) -- (345.7,235.11) .. controls (345.19,235.11) and (344.79,234.71) .. (344.79,234.2) -- cycle ;
%Straight Lines [id:da942833508902559] 
\draw    (355.41,186.09) -- (355.57,233.31) ;
%Shape: Rectangle [id:dp42131537768393246] 
\draw  [fill={rgb, 255:red, 255; green, 255; blue, 255 }  ,fill opacity=1 ] (351.85,194.2) -- (359.29,194.2) -- (359.29,201.59) -- (351.85,201.59) -- cycle ;
%Shape: Rectangle [id:dp24504883784733789] 
\draw  [fill={rgb, 255:red, 255; green, 255; blue, 255 }  ,fill opacity=1 ] (351.53,220.5) -- (358.96,220.5) -- (358.96,227.89) -- (351.53,227.89) -- cycle ;
%Rounded Rect [id:dp6906180993823305] 
\draw  [fill={rgb, 255:red, 0; green, 0; blue, 0 }  ,fill opacity=1 ] (430.77,231.69) .. controls (430.77,231.25) and (431.13,230.9) .. (431.56,230.9) -- (488.87,230.9) .. controls (489.31,230.9) and (489.66,231.25) .. (489.66,231.69) -- (489.66,234.08) .. controls (489.66,234.52) and (489.31,234.87) .. (488.87,234.87) -- (431.56,234.87) .. controls (431.13,234.87) and (430.77,234.52) .. (430.77,234.08) -- cycle ;
%Straight Lines [id:da8749672154064432] 
\draw    (460.22,232.88) -- (460.77,289.43) ;
%Shape: Rectangle [id:dp15567362116069416] 
\draw  [fill={rgb, 255:red, 255; green, 255; blue, 255 }  ,fill opacity=1 ] (456.45,239.64) -- (463.88,239.64) -- (463.88,247.03) -- (456.45,247.03) -- cycle ;
%Shape: Rectangle [id:dp6252244813004042] 
\draw  [fill={rgb, 255:red, 255; green, 255; blue, 255 }  ,fill opacity=1 ] (457.03,277.04) -- (464.47,277.04) -- (464.47,284.43) -- (457.03,284.43) -- cycle ;
%Rounded Rect [id:dp09133906240818024] 
\draw  [fill={rgb, 255:red, 0; green, 0; blue, 0 }  ,fill opacity=1 ] (500.65,231.63) .. controls (500.65,231.14) and (501.05,230.74) .. (501.54,230.74) -- (548.8,230.74) .. controls (549.29,230.74) and (549.69,231.14) .. (549.69,231.63) -- (549.69,234.28) .. controls (549.69,234.77) and (549.29,235.16) .. (548.8,235.16) -- (501.54,235.16) .. controls (501.05,235.16) and (500.65,234.77) .. (500.65,234.28) -- cycle ;
%Straight Lines [id:da8928039147505256] 
\draw    (520.26,186.09) -- (520.42,233.31) ;
%Shape: Rectangle [id:dp8817138301012912] 
\draw  [fill={rgb, 255:red, 255; green, 255; blue, 255 }  ,fill opacity=1 ] (516.7,194.2) -- (524.14,194.2) -- (524.14,201.59) -- (516.7,201.59) -- cycle ;
%Shape: Rectangle [id:dp8495071959923477] 
\draw  [fill={rgb, 255:red, 255; green, 255; blue, 255 }  ,fill opacity=1 ] (516.38,220.5) -- (523.81,220.5) -- (523.81,227.89) -- (516.38,227.89) -- cycle ;
%Rounded Rect [id:dp573910339935265] 
\draw  [fill={rgb, 255:red, 0; green, 0; blue, 0 }  ,fill opacity=1 ] (497.94,289.07) .. controls (497.94,288.58) and (498.33,288.19) .. (498.82,288.19) -- (546.09,288.19) .. controls (546.58,288.19) and (546.97,288.58) .. (546.97,289.07) -- (546.97,291.72) .. controls (546.97,292.21) and (546.58,292.61) .. (546.09,292.61) -- (498.82,292.61) .. controls (498.33,292.61) and (497.94,292.21) .. (497.94,291.72) -- cycle ;
%Straight Lines [id:da4462242315625988] 
\draw    (520.06,232.74) -- (520.53,290.77) ;
%Shape: Rectangle [id:dp13179931378426768] 
\draw  [fill={rgb, 255:red, 255; green, 255; blue, 255 }  ,fill opacity=1 ] (516.51,240.85) -- (523.94,240.85) -- (523.94,248.24) -- (516.51,248.24) -- cycle ;
%Shape: Rectangle [id:dp6995030495251187] 
\draw  [fill={rgb, 255:red, 255; green, 255; blue, 255 }  ,fill opacity=1 ] (516.96,277.17) -- (524.39,277.17) -- (524.39,284.56) -- (516.96,284.56) -- cycle ;
%Straight Lines [id:da0580974791639326] 
\draw    (447.12,186.21) -- (447.28,233.43) ;
%Shape: Rectangle [id:dp8774533929587145] 
\draw  [fill={rgb, 255:red, 255; green, 255; blue, 255 }  ,fill opacity=1 ] (443.97,193.38) -- (451.41,193.38) -- (451.41,200.77) -- (443.97,200.77) -- cycle ;
%Shape: Rectangle [id:dp17449492897039565] 
\draw  [fill={rgb, 255:red, 255; green, 255; blue, 255 }  ,fill opacity=1 ] (443.59,219.02) -- (451.02,219.02) -- (451.02,226.41) -- (443.59,226.41) -- cycle ;
%Rounded Rect [id:dp7657060013583292] 
\draw  [fill={rgb, 255:red, 0; green, 0; blue, 0 }  ,fill opacity=1 ] (436.25,288.1) .. controls (436.25,287.61) and (436.64,287.22) .. (437.13,287.22) -- (484.4,287.22) .. controls (484.89,287.22) and (485.28,287.61) .. (485.28,288.1) -- (485.28,290.75) .. controls (485.28,291.24) and (484.89,291.63) .. (484.4,291.63) -- (437.13,291.63) .. controls (436.64,291.63) and (436.25,291.24) .. (436.25,290.75) -- cycle ;
%Straight Lines [id:da2050782885238931] 
\draw    (406.66,188.41) -- (407.24,287.37) ;
%Rounded Rect [id:dp6980770807857435] 
\draw  [fill={rgb, 255:red, 0; green, 0; blue, 0 }  ,fill opacity=1 ] (349.05,286.31) .. controls (349.05,285.85) and (349.42,285.48) .. (349.88,285.48) -- (428.99,285.48) .. controls (429.44,285.48) and (429.82,285.85) .. (429.82,286.31) -- (429.82,288.79) .. controls (429.82,289.25) and (429.44,289.62) .. (428.99,289.62) -- (349.88,289.62) .. controls (349.42,289.62) and (349.05,289.25) .. (349.05,288.79) -- cycle ;
%Rounded Rect [id:dp1034902987850177] 
\draw  [fill={rgb, 255:red, 0; green, 0; blue, 0 }  ,fill opacity=1 ] (338.71,334.76) .. controls (338.71,334.27) and (339.11,333.87) .. (339.59,333.87) -- (386.86,333.87) .. controls (387.35,333.87) and (387.75,334.27) .. (387.75,334.76) -- (387.75,337.41) .. controls (387.75,337.9) and (387.35,338.29) .. (386.86,338.29) -- (339.59,338.29) .. controls (339.11,338.29) and (338.71,337.9) .. (338.71,337.41) -- cycle ;
%Straight Lines [id:da10695032414596617] 
\draw    (358.32,289.22) -- (358.48,336.44) ;
%Shape: Rectangle [id:dp380260869990777] 
\draw  [fill={rgb, 255:red, 255; green, 255; blue, 255 }  ,fill opacity=1 ] (354.76,297.33) -- (362.19,297.33) -- (362.19,304.72) -- (354.76,304.72) -- cycle ;
%Shape: Rectangle [id:dp45734573117351407] 
\draw  [fill={rgb, 255:red, 255; green, 255; blue, 255 }  ,fill opacity=1 ] (354.43,323.63) -- (361.87,323.63) -- (361.87,331.02) -- (354.43,331.02) -- cycle ;
%Rounded Rect [id:dp3359039135714026] 
\draw  [fill={rgb, 255:red, 0; green, 0; blue, 0 }  ,fill opacity=1 ] (336,392.2) .. controls (336,391.72) and (336.39,391.32) .. (336.88,391.32) -- (384.15,391.32) .. controls (384.64,391.32) and (385.03,391.72) .. (385.03,392.2) -- (385.03,394.85) .. controls (385.03,395.34) and (384.64,395.74) .. (384.15,395.74) -- (336.88,395.74) .. controls (336.39,395.74) and (336,395.34) .. (336,394.85) -- cycle ;
%Straight Lines [id:da5483687727403721] 
\draw    (358.12,335.87) -- (358.59,393.91) ;
%Shape: Rectangle [id:dp12178120328696251] 
\draw  [fill={rgb, 255:red, 255; green, 255; blue, 255 }  ,fill opacity=1 ] (354.57,343.98) -- (362,343.98) -- (362,351.37) -- (354.57,351.37) -- cycle ;
%Shape: Rectangle [id:dp7637358087969361] 
\draw  [fill={rgb, 255:red, 255; green, 255; blue, 255 }  ,fill opacity=1 ] (355.02,380.3) -- (362.45,380.3) -- (362.45,387.69) -- (355.02,387.69) -- cycle ;
%Rounded Rect [id:dp9651311313595947] 
\draw  [fill={rgb, 255:red, 0; green, 0; blue, 0 }  ,fill opacity=1 ] (407.75,345.94) .. controls (407.75,345.45) and (408.15,345.06) .. (408.64,345.06) -- (455.91,345.06) .. controls (456.39,345.06) and (456.79,345.45) .. (456.79,345.94) -- (456.79,348.59) .. controls (456.79,349.08) and (456.39,349.47) .. (455.91,349.47) -- (408.64,349.47) .. controls (408.15,349.47) and (407.75,349.08) .. (407.75,348.59) -- cycle ;
%Straight Lines [id:da575386374944223] 
\draw    (429.88,289.6) -- (430.35,347.64) ;
%Shape: Rectangle [id:dp6212164141878413] 
\draw  [fill={rgb, 255:red, 255; green, 255; blue, 255 }  ,fill opacity=1 ] (426.32,297.71) -- (433.76,297.71) -- (433.76,305.1) -- (426.32,305.1) -- cycle ;
%Shape: Rectangle [id:dp9198497418543585] 
\draw  [fill={rgb, 255:red, 255; green, 255; blue, 255 }  ,fill opacity=1 ] (426.77,334.04) -- (434.21,334.04) -- (434.21,341.43) -- (426.77,341.43) -- cycle ;
%Rounded Rect [id:dp3637640672218845] 
\draw  [fill={rgb, 255:red, 0; green, 0; blue, 0 }  ,fill opacity=1 ] (336.97,451) .. controls (336.97,450.51) and (337.36,450.11) .. (337.85,450.11) -- (385.12,450.11) .. controls (385.6,450.11) and (386,450.51) .. (386,451) -- (386,453.65) .. controls (386,454.14) and (385.6,454.53) .. (385.12,454.53) -- (337.85,454.53) .. controls (337.36,454.53) and (336.97,454.14) .. (336.97,453.65) -- cycle ;
%Straight Lines [id:da38386437977019416] 
\draw    (359.09,394.66) -- (359.56,452.7) ;
%Shape: Rectangle [id:dp3399274835213124] 
\draw  [fill={rgb, 255:red, 255; green, 255; blue, 255 }  ,fill opacity=1 ] (355.54,402.77) -- (362.97,402.77) -- (362.97,410.16) -- (355.54,410.16) -- cycle ;
%Shape: Rectangle [id:dp20767760296409654] 
\draw  [fill={rgb, 255:red, 255; green, 255; blue, 255 }  ,fill opacity=1 ] (355.99,439.1) -- (363.42,439.1) -- (363.42,446.49) -- (355.99,446.49) -- cycle ;
%Shape: Rectangle [id:dp7490862728828036] 
\draw  [line width=0.75]  (384.18,9.95) -- (412.13,9.95) -- (412.13,23.93) -- (384.18,23.93) -- cycle ;
%Shape: Diamond [id:dp8784364359623389] 
\draw  [line width=0.75]  (398.48,9.87) -- (411.05,16.9) -- (398.48,23.93) -- (385.9,16.9) -- cycle ;
%Straight Lines [id:da8663411815883146] 
\draw [line width=0.75]    (384.18,9.95) -- (412.13,23.93) ;
%Straight Lines [id:da5710863316747317] 
\draw [line width=0.75]    (384.18,23.93) -- (412.13,9.95) ;
%Straight Lines [id:da36900576552002473] 
\draw    (398.07,23.93) -- (398.05,38.88) ;
%Shape: Rectangle [id:dp45935186249683757] 
\draw  [line width=0.75]  (345.84,467.02) -- (373.78,467.02) -- (373.78,481) -- (345.84,481) -- cycle ;
%Shape: Diamond [id:dp6918367395746916] 
\draw  [line width=0.75]  (360.13,466.93) -- (372.71,473.97) -- (360.13,481) -- (347.56,473.97) -- cycle ;
%Straight Lines [id:da013987991524053278] 
\draw [line width=0.75]    (345.84,467.02) -- (373.78,481) ;
%Straight Lines [id:da2867633085143759] 
\draw [line width=0.75]    (345.84,481) -- (373.78,467.02) ;
%Straight Lines [id:da6874439590639045] 
\draw    (359.56,452.7) -- (359.54,467.65) ;

% Text Node
\draw (375.85,32.35) node   [align=left] {1};
% Text Node
\draw (351.12,78.73) node   [align=left] {2};
% Text Node
\draw (338.55,127.37) node   [align=left] {4};
% Text Node
\draw (433.9,128.98) node   [align=left] {3};
% Text Node
\draw (396.7,178.42) node   [align=left] {8};

% Text Node
\draw (312.17,177.43) node   [align=left] {5};
% Text Node
\draw (281.47,222.08) node   [align=left] {6};
% Text Node
\draw (343.62,222.76) node   [align=left] {7};
% Text Node
\draw (438.24,222.98) node   [align=left] {9};
% Text Node
\draw (443.41,280.43) node   [align=left] {10};
% Text Node
\draw (508.7,223.05) node   [align=left] {11};
% Text Node
\draw (505.6,280.11) node   [align=left] {12};
% Text Node
\draw (344.76,326.18) node   [align=left] {13};
% Text Node
\draw (343.66,383.24) node   [align=left] {15};
% Text Node
\draw (361.31,277.02) node   [align=left] {14};
% Text Node
\draw (415.42,336.97) node   [align=left] {16};
% Text Node
\draw (345.63,442.03) node   [align=left] {17};
% Text Node
\draw (414,5) node [anchor=north west][inner sep=0.75pt]   [align=left] {{\scriptsize $S_1$}};
% Text Node
\draw (376,464) node [anchor=north west][inner sep=0.75pt]   [align=left] {{\scriptsize $S_1$}};

\draw  [fill={rgb, 255:red, 255; green, 255; blue, 255 }  ,fill opacity=1 ] (402.7,192.42) -- (410.13,192.42) -- (410.13,200.42) -- (402.7,200.42) -- cycle ;

\draw  [fill={rgb, 255:red, 255; green, 255; blue, 255 }  ,fill opacity=1 ] (403.7,274.42) -- (411.13,274.42) -- (411.13,282.42) -- (403.7,282.42) -- cycle ;

\end{tikzpicture}

  \end{minipage}
  %\hfill
  \begin{varwidth}[b]{0.25\linewidth}
  {
      \renewcommand{\arraystretch}{1.2}
      \centering
      \begin{tabular}{c|c}
          Bus id & $P_f$ \\
          \hline
            1 & 0.125 \\
            2 & 0.500 \\ 
            3 & 0.250 \\ 
            4 & 0.500 \\ 
            5 & 0.500 \\ 
            6 & 0.500 \\ 
            7 & 0.125 \\ 
            8 & 0.125 \\ 
            9 & 0.125 \\
            10 & 0.500 \\ 
            11 & 0.250 \\ 
            12 & 0.500 \\ 
            13 & 0.500 \\ 
            14 & 0.500 \\ 
            15 & 0.125 \\ 
            16 & 0.125 \\
            17 & 0.125 \\
            \multicolumn{1}{c}{} \\
      \end{tabular}
    %\color{white}
    %\caption{
    %Probability of Failure
    %}
\hspace{0pt}
\vfill
  }
  \end{varwidth}%
  \vspace{8pt}
    \captionof{figure}{17-bus distribution system and the probability of failure ($P_f$) values for each bus.}
    \label{fig:mid_size}
\end{table}